\documentclass[final,5p,times,twocolumn]{elsarticle}

\usepackage{amssymb}
\usepackage{gensymb}
\usepackage{amsmath}
\usepackage{amsthm}
\usepackage{textcomp}
\usepackage{hyperref}
\usepackage{afterpage}
\usepackage{comment}
\usepackage{graphicx} % for placing figures
\usepackage{subfigure}
\usepackage{float} % For allowing graphics, and other objects, to float around the page
\usepackage{url}
\usepackage{xcolor} %or {xcolor}. % Allow text to be colored
\usepackage{enumitem} % for enumerating items
\usepackage{adjustbox}
\usepackage{titlesec}
\usepackage{stfloats}
\usepackage{setspace}
\usepackage{multirow} % multiple rows in tables
\usepackage{siunitx} % SI units for consistency
\usepackage{textcomp}
\usepackage[version=4]{mhchem} % chemical equations
\usepackage[capitalise, nameinlink, noabbrev]{cleveref}
\usepackage{soul}
\usepackage[utf8]{inputenc}
\usepackage{array}

\usepackage{booktabs}
\hypersetup{
    colorlinks=true,
    linkcolor=cyan,
    filecolor=magenta,      
    urlcolor=blue
    } 
\usepackage{etoolbox}
\usepackage{lineno}
\journal{Combustion and Flame}

\begin{document}

\newpage
\begin{frontmatter}

\title{Machine Learning-Assisted Analysis of Combustion and Ignition in As-milled and Annealed Al/Zr Composite Powders}

%\title{CNN Assisted Investigation of Ignition and Combustion in Al/Zr Composite Powders: The Role of Intermetallic Heat}

\author[inst1,inst2]{Michael R. Flickinger\corref{cor1}}

\affiliation[inst1]{organization={Department of Materials Science and Engineering, Johns Hopkins University},%Department and Organization
            addressline={3400 N Charles St}, 
            city={Baltimore},
            postcode={21218}, 
            state={MD},
            country={U.S.A}}

\affiliation[inst2]{organization={Hopkins Extreme Materials Institute, Johns Hopkins University},%Department and Organization
            addressline={3400 N Charles St}, 
            city={Baltimore},
            postcode={21218}, 
            state={MD},
            country={U.S.A}}

\ead{mflicki1@jhu.edu}
\cortext[cor1]{Corresponding authors}

\author[inst1,inst2]{Sreenivas Raguraman}

\author[inst1,inst3]{Amee L. Polk}
\affiliation[inst3]{organization={Department of the Army, U.S. Army Combat Capabilities Development Command Chemical Biological Center},
            addressline={8188 Blackhawk Road}, 
            city={Aberdeen Proving Ground},
            postcode={21010}, 
            state={MD},
            country={U.S.A}}

\author[inst2,inst4]{Colin Goodman}

\author [inst1,inst2]{Megan Bokhoor}

\author[inst1]{Rami Knio}

\author [inst2,inst4]{Michael Kruppa}

\author[inst2,inst4]{Mark A. Foster}

\affiliation[inst4]{organization={Department of Electrical and Computer Engineering, Johns Hopkins University},
            addressline={3400 N Charles St}, 
            city={Baltimore},
            postcode={21218}, 
            state={MD},
            country={U.S.A}}

\affiliation[inst5]{organization={Department of Mechanical Engineering, Johns Hopkins University},
            addressline={3400 N Charles St}, 
            city={Baltimore},
            postcode={21218}, 
            state={MD},
            country={U.S.A}}

\author[inst1,inst2,inst5]{Timothy P. Weihs\corref{cor1}}
\ead{weihs@jhu.edu}
%\cortext[cor2]{Corresponding authors}

\begin{abstract}

Micron-scale metal-based composite powders are promising for energetic applications due to their tailored ignition and combustion properties. In particular, ball-milled Al/Zr composites exhibit lower ignition thresholds than pure aluminum, driven by exothermic intermetallic formation reactions and have demonstrated enhanced combustion properties. However, the extent to which this heat release governs ignition and combustion remains unclear, especially when progressively removed through annealing. To systematically investigate this effect, we synthesized Al/Zr powders (3Al:Zr, Al:Zr, and Al:3Zr at\%) via ball milling, annealed them in argon up to 1000 °C to partially complete the formation reactions, and characterized their ignition and combustion behavior. Ignition thresholds were measured using a hot wire method across different environments, while high-speed hyperspectral imaging tracked single-particle burn durations and temperatures. A convolutional neural network (CNN)-based method was developed to quantify the frequency of microexplosions. Results show that annealing—and thus reducing available reaction heat—increases ignition thresholds, most significantly for Al-rich compositions. In contrast, Zr-rich powders exhibit little change in ignition thresholds due to oxidation aiding ignition. Despite removing the available heat that drives ignition, average combustion temperatures range from ~2400 K to ~3000 K and increased with annealing for Al- and Zr-rich powders. Average maximum temperatures are 100 to 400 K higher. The frequency of microexplosions remains high (>46\%) and increases with annealing for all but the Al-rich powders. These findings suggest that while homogeneous Al/Zr powders (e.g., atomized) may exhibit higher ignition thresholds, they can achieve comparable combustion performance once ignited.  

\end{abstract}


\end{frontmatter}

\section{Introduction}
\label{sec:Introduction}

% General overview kind of statements to set the stage
Metal-based powders are commonly used in explosive, propellant, pyrotechnic, and biological/chemical agent defeat applications due to their high heats of combustion \cite{yen2012reactive, peiris2019applications,camilleri2020mechanisms, dreizin2000phase, grinshpun2017aluminum, grinshpun2012inactivation, borah2024development}. Among the metal-based powders, micron-sized aluminum (Al) is the most frequently used metal additive for energetics due to its low cost, relatively high energy density, and ease of handling \cite{dreizin2009metal, sundaram2016general}. Nonetheless, the complete benefits of micron-sized Al powder have yet to be actualized due to its passivating native oxide, which leads to ignition delays and suboptimal combustion performance \cite{sundaram2016general}.

% ball-milling of composite metal powders for lowering and tuning ignition thresholds
Researchers have improved the ignition and combustion performance of pure Al by mixing it with other metals that have high heats of combustion, such as B, Mg, Si, Ti, and Zr, to form composite or alloy powders \cite{shoshin2006particle, wainwright2023comparing, feng2020ignition, ma2022spherical, zhu2007mechanically, le2023situ, herbold2011effects}. In particular, the creation of micron-sized composite powders through mechanical ball-milling is known to drastically reduce ignition thresholds \cite{dreizin2017mechanochemically, suryanarayana2019mechanical} compared to pure Al powder, which ignites anywhere from aluminum's melting point ($\sim$660 °C) for nanoscale powders to alumina's melting temperature ($\sim$2000 °C) for microscale powders \cite{dreizin2009metal}. Binary composite metal powders exhibit lower ignition thresholds due to exothermic reactions between their two metallic components if they have a high negative heat of mixing \cite{dreizin2017mechanochemically,lakshman2019milling}. 

Besides chemistry, ignition thresholds can also be tuned by varying milling parameters such as the milling time, process control agent (PCA), milling media, and ball-to-powder ratio (BPR) \cite{lakshman2019milling, lakshman2021evaluating}. Changing powder chemistry and milling conditions alter reactant spacings and degrees of premixing, which in turn strongly influence the ignition thresholds \cite{arlington2022exothermic,lakshman2019milling, lakshman2021evaluating, wainwright2023comparing}. Over-milling of the metal reactants, for example, raises the ignition threshold as a smaller fraction of the intermetallic heat of reaction is available to drive ignition \cite{lakshman2019milling}.

% ball-milling of composite metal powders for enhancing combustion properties
The ball-milled composite powders have also demonstrated improved combustion performance. Explosive launch experiments have shown that Al/Mg/Zr composite powders have significantly higher combustion efficiencies than 10$\times$ smaller Al powders \cite{stamatis2020combustion}. Composite, ball-milled particles can extend their burn duration through a dual-phase combustion mechanism whereby one element, such as Al, burns in the vapor phase while the other element burns in the condensed phase (e.g., Zr, Ti) \cite{wainwright2019bubbling, wainwright2023comparing, steinberg1992combustion}. Moreover, these composite particles can microexplode, causing the molten particles to violently fragment into smaller burning fragments \cite{mcnanna2025disruptive}. Microexplosion fragmentation is known to enhance combustion efficiency in mixed fuel droplets \cite{nyashina2020impact, meng2021experimental, wang2025effect, zhou2021effect}, and is thought to improve the combustion efficiency of metal powders by rapidly exposing new surfaces to the oxidizer. 

Despite recent advances in understanding how powder chemistry and milling conditions impact the ignition thresholds of ball-milled powders, little work has been done to determine how varying the heat from intermetallic formation reactions for a given microstructure impacts the ignition and combustion properties of the powders. A study by Aly \textit{et al.} \cite{aly2015ignition} examined the ignition and combustion properties of two Al:Mg powders of the same chemistry fabricated by either mechanical milling or by casting and crushing an intermetallic alloy. Mechanically milled Al:Mg powders were found to have a slightly lower ignition threshold and a two-stage burn mechanism, unlike Al:Mg powders made by casting and crushing \cite{aly2015ignition}. However, their study only recorded the average combustion temperatures for the mechanically milled Al:Mg samples and thus did not compare them to the temperatures of the cast and crushed powders. 

In this study, we investigate the influence of intermetallic heats of formation in three stoichiometric ball-milled Al/Zr powders (3Al:Zr, Al:Zr, and Al:3Zr at\%) by systematically annealing the powders by slowly heating them in argon (Ar) to increasingly higher temperatures, thus progressively reducing the available heat of reaction for a given initial microstructure. The Al/Zr system was chosen as it has been studied extensively previously \cite{wainwright2018gasenvs, wainwright2019bubbling, wainwright2020microstructure,ma2022spherical}. We perform thermal analysis at slow heating rates to both reduce and evaluate the powders' available heats of reaction following annealing, and to characterize their oxidation/nitridation behavior. 

We quantify ignition thresholds of the annealed samples using hot wire ignition \cite{ward2006wireignition}, and use a custom built high-speed hyperspectral imaging diagnostic tool known as Snapshot Hyperspectral imager for Emission And Reactions (SHEAR) to capture the as-milled and annealed powder's combustion behavior in a high-throughput manner \cite{alemohammad2020kilohertz}. Hyperspectral imaging analyzes a particle's spectrum across hundreds of wavelengths, rather than relying on the ratio of just two or three wavelengths, as in two- or three-color pyrometry, and has been used to measure temperatures of flames and single particles \cite{alemohammad2020kilohertz,hameete2024particle, peng2023combustion,goroshin2007emission,si2021study}. To quantify a particle's temperature throughout its burn duration, this study develops a tailored multi-object tracking framework augmented by machine learning corrections to simultaneously track numerous particles within the field of view. 

Finally, while some studies have attempted to quantify the frequency of microexplosions in single-metal combustion experiments for specific powder chemistries and combustion conditions \cite{huang2022detailed, zhou2025fundamental, peng2025micro}, quantitative results reporting these values remain limited. Moreover, these efforts are often hindered by the need to manually identify microexplosion events, which can be very inefficient when the amount of data required for statistical conclusions is substantial. To address this, we developed a set of three different machine learning based convolutional neural networks (CNN) using deep learning techniques to analyze small regions surrounding particles' last known locations and evaluate whether a microexplosion occurred. 
%The methodology and training of these networks are described in the next section. 

%===============================================================================

\section{Materials and Methods}
\label{sec:Materials}

%------------------------------------------------------------------------------------------------------
\subsection{Powder Synthesis \& Annealing}
\label{subsec:snythesis}

Three different powder chemistries were synthesized: 3Al:Zr, Al:Zr, and Al:3Zr (atomic ratio) in a SPEX 8000D shaker mill using commercial Al (Alfa Aesar, 99.5 \% pure, -325 mesh) and Zr (Atlantic Equipment Engineers, 99.5 \% pure, -20 to +60 mesh) elemental powders. For each chemistry, around 10 grams of powder were placed into a hardened steel vial with 10 mL of hexane as a process control agent and 100 grams of 3/8" stainless steel balls. The contents were then placed in the shaker mill and milled for one hour. The powders were subsequently sieved for 3h to <75 {\textmu}m using a motorized sieve shaker (Dual Manufacturing Co. Inc., D-4326) in a 3” diameter brass sieve with a pore size of 75 {\textmu}m. Based on previous studies, we presume that the average composition of Al/Zr ball-milled powders is not altered significantly by sieving \cite{wainwright2020microstructure}.

The three powders were scanned to different temperatures in Ar inside a Differential Thermal Analysis (DTA) up to 1000 °C to reduce the intermetallic heat of formation that is available for release. Here, we refer to this process as annealing, although traditionally annealing is performed at a constant temperature. In this work, we use the term more broadly to refer to heating the powders at a constant rate to a target temperature and subsequently air cooling to room temperature. 

Samples were heated in 90 $\mu$L alumina crucibles in Ar (99.999\%) at 20 °C/min to the target temperatures shown in Table~\ref{tab:sample_table}. The Al:3Zr sample was heated to an additional temperature prior to the onset of its main endothermic peak. The amount of intermetallic heat released on scanning to 1000 °C, $\Delta H_{1000C}$, was calculated by subtracting a second, baseline scan, from the first scan and integrating from 150 °C to 1000 °C. The heat released on scanning to lower temperatures was calculated in a similar manner, and the percentage of heat released up to that temperature, $P_{T}$, is given by:

\begin{equation}\label{eqn:Heat_removed}
    P_T = 100 \times \dfrac{\Delta H_T}{\Delta H_{1000\degree \text{C}}} 
\end{equation}

The maximum temperatures to which samples were heated and the percentages of intermetallic heat removed for each sample are shown in \cref{tab:sample_table}. Starting ball milled powders are referred to as as-milled, and annealed powders are labeled by their composition, maximum scan temperature, and  $P_{T}$ in parentheses (e.g., 3Al:Zr 430 °C (25\%)). 

\begin{table}[h]
\centering
\begin{tabular}{ccc}
\toprule
\begin{tabular}[c]{@{}c@{}}Base Composition\\ (atomic ratio)\end{tabular} & Anneal Temperature ( °C) & \begin{tabular}[c]{@{}c@{}} $P_T$ (\%)\end{tabular} \\
\midrule
\multirow{5}{*}{3Al:Zr} & 0   & 0     \\
                        & 430 & 25  \\
                        & 500 & 49  \\
                        & 590 & 82  \\
                        & 1000 & 100  \\
\midrule
\multirow{5}{*}{Al:Zr}  & 0   & 0     \\
                        & 350 & 21  \\
                        & 490 & 50  \\
                        & 640 & 76  \\
                        & 1000 & 100  \\
\midrule
\multirow{6}{*}{Al:3Zr} & 0   & 0     \\
                        & 360 & 14  \\
                        & 470 & 36  \\
                        & 550 & 55  \\
                        & 750 & 100   \\
                        & 1000 & 100* \\
\bottomrule
\end{tabular}
\caption{List of as-milled and annealed samples used in this study with their respective chemistry, maximum scan temperature, and amount of intermetallic heat removed. $P_T$ is the percentage of intermetallic heat removed calculated using \cref{eqn:Heat_removed} rounded to the nearest whole number to simplify the sample naming. The asterisks next to 100\% of intermetallic heat removed for the Al:3Zr 1000 °C sample denote that after 750 °C, there is a large endotherm present to 1000 °C, which is not related to the exothermic heat from the intermetallic formation reactions.}
\label{tab:sample_table}
\end{table}

%------------------------------------------------------------------------------------------------------
\subsection{Powder Characterization}
\label{subsec:characterization}

Particle size was determined using a Horiba LA-950 V2 particle size analyzer. Powders were dispersed in isopropanol and agitated to break up agglomerates. Six measurements were averaged to obtain the reported particle size distributions. The sieved as-milled powders were placed onto aluminum stubs with carbon tape and imaged using a scanning electron microscope (Tescan MIRA 3 GM field-emission SEM) to characterize powder size and morphology. As-milled powders were, in addition, mounted in epoxy and polished for cross-sectional SEM imaging with a backscatter detector to observe the distribution and morphology of Zr inclusions in the Al composite matrix using the JEOL IT700HR InTouchScopeTM microscope.

\begin{figure*}[b!]
    \centering
    \includegraphics[width=.95\linewidth]{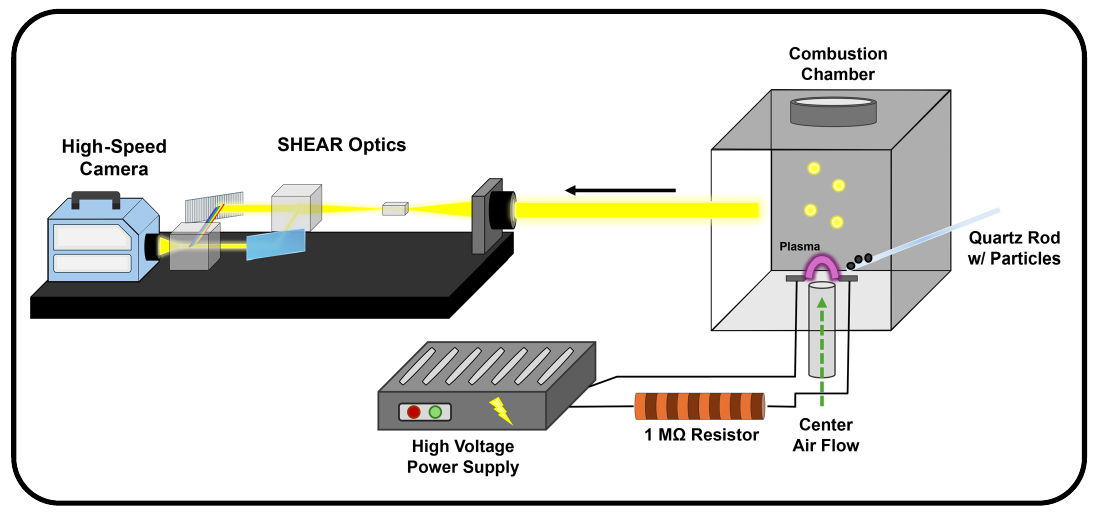}
    \caption{A schematic showing the experimental setup used for combustion experiments. Note that the optics setup shown is not meant to be an exact representation of the optics used in SHEAR.}
    \label{shear_schematic}
\end{figure*}

A TA Instruments Discovery SDT 650 was used to perform Differential Thermal Analysis (DTA) and Thermogravimetric Analysis (TGA) on the as-milled powders to measure the heat of the intermetallic formation reactions and the mass gain in different gaseous environments. Powders were heated twice from 50 °C to 1000 °C at a rate of 20 °C/min under high-purity Ar flowing at 100 mL/min. The second scan was used as a baseline and was subtracted to obtain the net heat flow from the irreversible intermetallic formation reactions. The mass gains due to just oxidation and just nitridation were assessed by heating the powders at the same rate in Ar + \ce{O2} (21 vol\% \ce{O2}) and Ar + \ce{N2} (78 vol\% \ce{N2}), respectively, at 100 mL/min of gas flow. The powders were also heated while flowing compressed air at 100 mL/min to measure the mass gain in a mixed environment. Lastly, samples were heated to 1000 °C in high-purity Ar, cooled, and then heated again in compressed house air to assess the main gain in air for fully annealed particles. A minimum of three trials were carried out in each DTA experiment. 

X-ray diffraction (XRD) was performed on samples heated in the DTA to identify crystalline phases using a Malvern Panalytical Aeris XRD (PIXcel1D-Medipix3 detector, Cu K$\alpha$ radiation, 40kV, 7.5mA). The samples were scanned on a low-background Si disk from 20 to 75 degrees 2$\theta$, and the overall intensity was summed from three scans. Data processing, such as background subtraction, sample misalignment correction,  removal of K$\alpha_2$ peaks, peak identification and pattern matching, and profile fitting, was performed using HighScore Plus4 analysis software as per instructions from \cite{speakman_profile_2012,raguraman_machine_2024}. 

Wire ignition tests were performed using a modified hot wire setup constructed within a vacuum chamber designed to allow for environmental control similar to prior studies \cite{polk2024thermite,lakshman2019milling,ward2006wireignition}. Experiments were performed in air, Ar + O\textsubscript{2} (21 vol\% O\textsubscript{2}), and Ar + N\textsubscript{2} (78 vol\% N\textsubscript{2}). Five fill-purge cycles were performed using an Edwards nXDS10i dry scroll pump, and we estimate that after five fill-purge cycles the remaining \ce{N2} content was 0.009 (vol\%) and the remaining \ce{O2} content was 0.0024 (vol\%). Nichrome wires (28-gauge, approximately 4 inches in length) were cut and pre-tensioned before being mounted between two electrodes across which a 36V charge was pulsed to resistively heat the filament. Heating rates varied from approximately 15,000 to 30,000 K/s. Before pulsing the wires with current, they were coated with dispersed powder-hexane slurries. 

Videos were taken using a Photron FASTCAM SA-Z at 20,000 frames per second. The pulse was delivered using a LABVIEW program that synchronized the camera trigger pulse and two-color infrared pyrometer (Kieber 740-LO) readings, allowing the moment of first light to be correlated with the wire temperature using a linear fit. The emissivity of the pyrometer was set to 0.74 to match that of the Nichrome wire \cite{ward2006wireignition}. The data analysis was performed using a customized Python script.

The combustion properties of as-milled powders and ones scanned to 1000 °C were evaluated using a custom-designed combustion chamber, schematically illustrated in \cref{shear_schematic}. A plasma was generated to ignite flowing powders by applying 15 kV between two tungsten electrodes that were spaced 9 mm apart. Approximately 15 to 30 mg of loose powders were loaded onto the end of a thin quartz rod and then inserted into the plasma. The particles were carried into the field of view with compressed air flowing at 1 cubic foot per minute. 

A high-speed camera (Photron FASTCAM SA-Z), filming at 20,000 frames per second and with a shutter speed of 1/40,000 seconds, captured the combustion scene using a hyperspectral optics configuration called Snapshot Hyperspectral Imager for Emissions and Reactions (SHEAR). The SHEAR optics create an imaging and co-registered spectral channel whereby a single particle's emission spectra can be correlated and fit to estimate temperature as described by Alemohammed et al. \cite{alemohammad2020kilohertz}. 

\subsection{SHEAR Particle Tracking \& Microexplosion Detection}
\label{subsec:characterization}

A suite of custom Python scripts was developed to efficiently identify, track, and record the particle trajectories and emission data. While the foundational principles of the SHEAR system are detailed extensively in the original reference \cite{alemohammad2020kilohertz}, we briefly highlight some of the specific components of our implementation, particularly the particle tracking algorithms used in the software.

\begin{figure*}[h]
    \centering
    \includegraphics[width=0.9\linewidth]{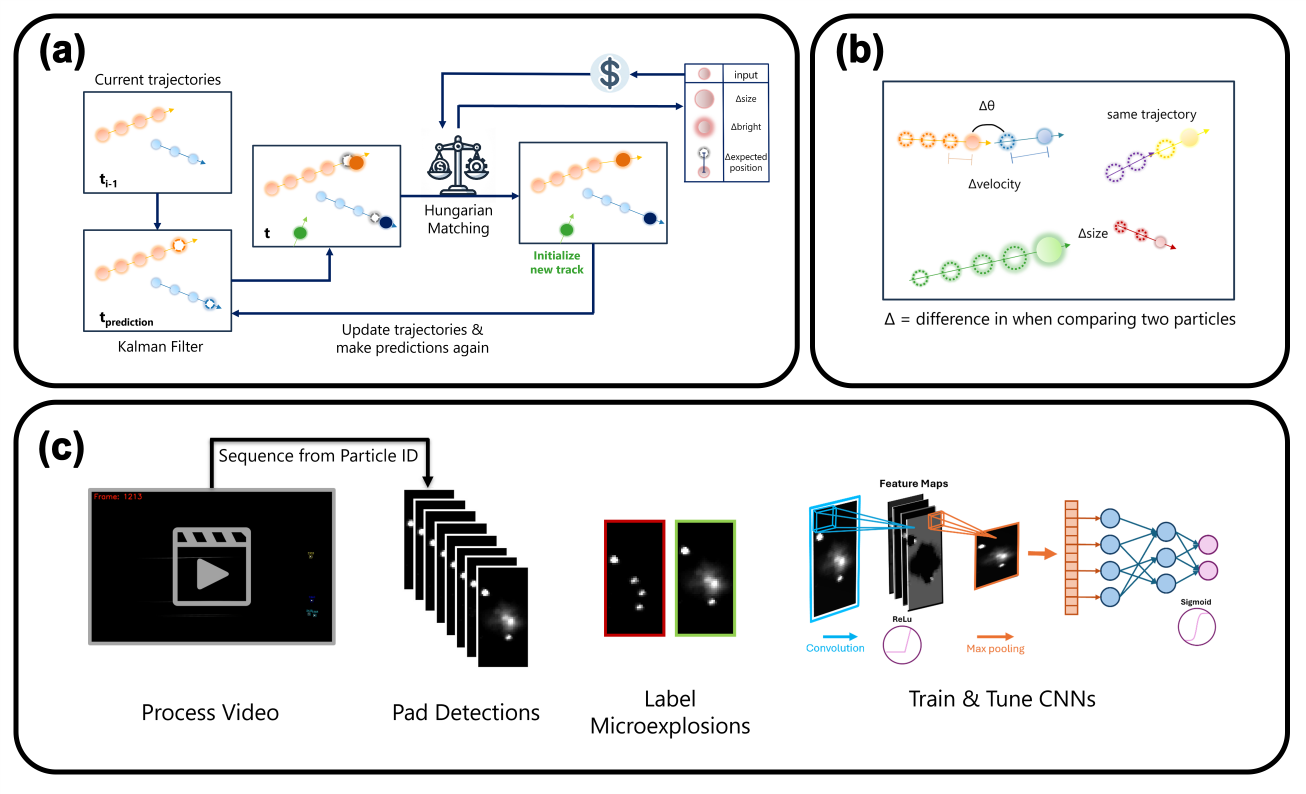}
    \caption{(a) A schematic representation of the Kalman filter and Hungarian Algorithm approach employed to track particles in SHEAR videos. (b) Shows the main features that are calculated and fed into a logistic regression classifier trained to merge particle trajectories that are likely of the same particle. (c) The general procedure used to create three different convolutional neural network (CNN) models was employed to predict whether particles microexploded in the SHEAR videos. Note that the schematic architecture shown in the image is not a precise representation of any of the three models that were trained, but rather encapsulates some general concepts commonly used in CNNs.}
    \label{kalman}
\end{figure*}

Combusting particles appear as small, bright spots in the videos, making detection a straightforward task. To detect them in each frame, median and Gaussian blurring were applied, followed by thresholding with OpenCV to generate a binary image \cite{opencv_library}. The detected particles were tracked using a Kalman Filter and  Hungarian Algorithm approach \cite{sahbani2016kalman,ciaparrone2020deep}. \cref{kalman} (a) schematically illustrates the tracking approach, where assignment costs are determined based on differences in bounding box size, local area brightness, and predicted position. Each factor contributes to an associated cost \( C \), which follows a piecewise function to enable smooth tracking while penalizing large mismatches:

\[
C_f(\Delta f) =
\begin{cases}
    w_f (\Delta f)^2, & |\Delta f| \leq f_{\text{thresh}} \\
    C_{\text{HIGH}}, & |\Delta f| > f_{\text{thresh}}
\end{cases}
\]

where \( \Delta f \) represents the difference in centroid position (\( d \)), bounding box size (\( s \)), or brightness (\( b \)), with respective weights \( w_d, w_s, w_b \).

The full assignment cost is then defined as:

\[
C_{i,j} = w_d (\Delta d)^2 + w_s (\Delta s)^2 + w_b (\Delta b)^2 + O
\]

where:
\begin{itemize}
    \item \( \Delta d \) is the Euclidean distance between predicted and observed centroids,
    \item \( \Delta s \) is the bounding box size difference,
    \item \( \Delta b \) is the absolute brightness difference.
\end{itemize}

Unlike the other cost terms, the occlusion penalty follows a cubic function to heavily penalize long-term occlusions:

\[
O = o_{\text{prev}}^3 + 10
\]

where \( o_{\text{prev}} \) is the number of previous occlusions.

A square cost matrix \( C \in \mathbb{R}^{K \times K} \) is then constructed, where \( K = \max(N, M) \) ensures all previous tracks and new detections are considered. If no valid assignment exists, zero-cost placeholders are used:

\[
C_{i,j} = 0, \quad \text{if } i \geq N \text{ or } j \geq M.
\]

The Hungarian algorithm \cite{sahbani2016kalman} is applied to minimize the total assignment cost:

\[
\min \sum_{i} C_{r_i, c_i}
\]

where \( r_i \) and \( c_i \) represent the optimal assignments.

To further refine assignments, the cost matrix is iteratively expanded if large costs remain. Additional rows and columns are assigned a large penalty:

\[
C_{n+1, i} = C_{\text{LARGE}}, \quad C_{i, n+1} = C_{\text{LARGE}}, \quad \forall i.
\]

This discourages spurious matches while improving overall assignment consistency.

Once optimal assignments are determined, particles are handled as follows:
\begin{itemize}
    \item If \( r_i \geq N \), a new particle is created.
    \item If \( c_i \geq M \), a particle is marked as occluded.
    \item Otherwise, the detected bounding box is assigned to the corresponding tracked particle.
\end{itemize}

This framework enables multi-object tracking by ensuring optimal assignments while handling occlusions and the appearance of new particles.

To further refine spurious ID switching, a machine learning-based tracking error correction was applied post-tracking. A logistic regression classifier was trained on an independently labeled dataset of Al/Mg/Zr powders and subsequently applied to the tracking results. \cref{kalman}(b) diagrams the features used in the model, which include size difference, centroid displacement, velocity difference, and angular difference. Over 650 ID switches were manually identified, forming the dataset used for training and testing. Following classification, particle identities were corrected using a connected component approach. A graph is then constructed, where each node represents a particle, and an edge exists between nodes classified as the same object. A union-find algorithm is applied to merge connected components, assigning the lowest ID within each group as the new particle ID.

\begin{figure*}[ht]
    \centering
    \includegraphics[width=0.7\linewidth]{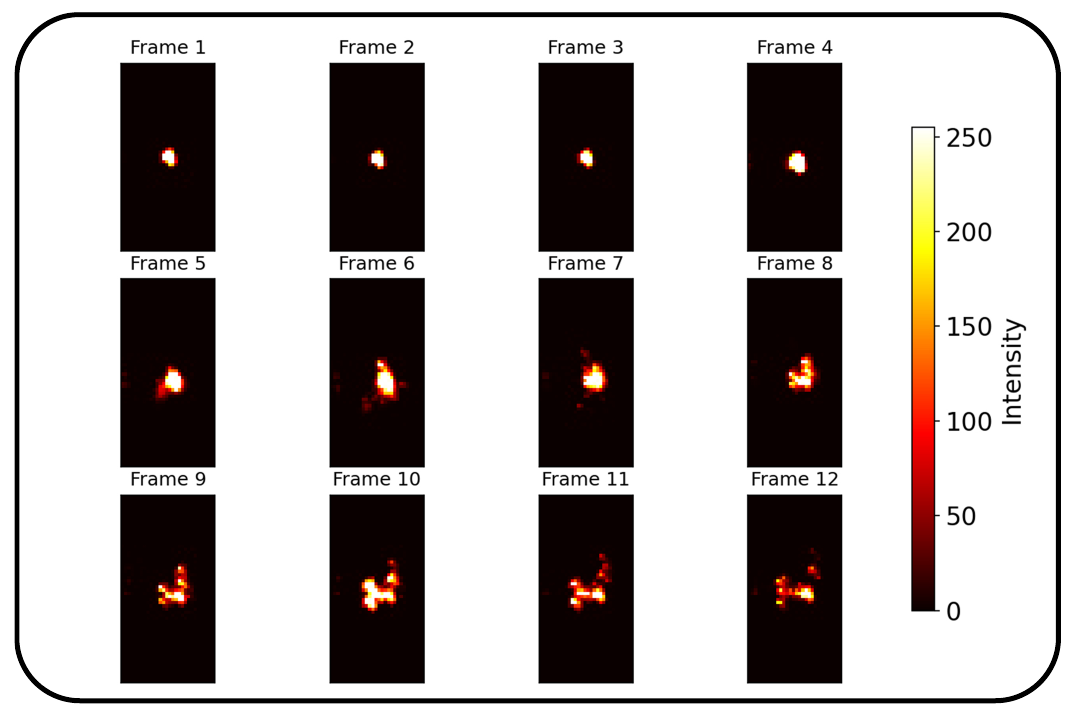}
    \caption{Frames before and after particle ID terminates due to the microexplosion of an Al:Zr as-milled particle. Frame 6 is the last frame before the particle spawns a new ID. A packet of frames like this was fed into a set of three CNN models to determine whether a microexplosion event occurred.}
    \label{me_frames}
\end{figure*}

Frequent microexplosions pose a significant problem for multi-object tracking of the combusting particles in the SHEAR videos. Typically, particles that microexplode have rapid jumps in apparent brightness and size, which are two of the main factors fed into the cost function for tracking. \cref{me_frames} displays video frames taken before and after a microexplosion. Particle fragmentation can rapidly generate multiple new IDs, and while these fragments are short-lived, they can disrupt the tracking algorithm by interfering with the occlusion handling process. Therefore, we developed a method for detecting microexplosions in SHEAR videos by looking at the local area around the particle six frames before and after its last known location. We chose to focus on a small padded region around each particle detection to account for the rapid expansion that occurs during the microexplosion, and to add additional context for classification. 

The general procedure for model training is shown in \cref{kalman}(c). Three CNNs were developed for binary microexplosion classification of the padded grayscale images (30 × 60 pixels). The models were: (1) VGG-like model \cite{simonyan2014very}, (2) LeNet-like model \cite{lecun1998gradient}, and (3) Simple CNN. The model architectures were chosen due to the relative simplicity of the images and the nature of the binary classification task. A more thorough explanation of the model architectures is provided in Supplementary Information Section S1.1. Examples of the padded grayscale images are shown in Supplementary Information Section S1.2.

A tunable focal loss function \cite{lin2017focal} was used to address the class imbalance (i.e., many more images were of the negative, not-microexplosion class), and a large dataset of over 10,000 images was used for training, testing, and validation. This dataset, comprising over 10,000 images, although larger than necessary for the immediate study, was utilized to develop a versatile tool capable of classifying microexplosion events across various powder chemistries and filming conditions, thereby ensuring its applicability to a broader range of combustion video analyses. All models were trained using the Adam optimizer \cite{kingma2014adam} with a tunable learning rate, determined through Bayesian optimization with Optuna \cite{akiba2019optuna}. 

Data augmentation techniques, including random rotations, width/height shifts, brightness changes, and horizontal flips, were applied to the training set. Early stopping and learning rate reduction were employed to prevent overfitting. The models were evaluated on a test set using precision, recall, F1-score, and the area under the precision-recall curve. Predictions were analyzed across multiple probability thresholds to determine the optimal threshold for classification. Training was conducted on an NVIDIA GeForce RTX 4060 Ti using TensorFlow \cite{tensorflow2015-whitepaper} and Keras \cite{chollet2015keras} within an Anaconda environment. Optuna was used for automated hyperparameter tuning, with early pruning based on validation loss. The performance of the three models on different test sets of data is shown in Supplementary Information Section S1.3. Further details of the combustion test experimental procedure and data analysis can be found in Supplementary Information Section S1.4. 

%=============================================================================
\section{Results}
\label{sec:Results}

\subsection{Particle Size, Morphology, and Microstructure} 
\label{subsec:sizemorphmicro}

\begin{figure*}[h!]
    \centering
    \includegraphics[width=1\linewidth]{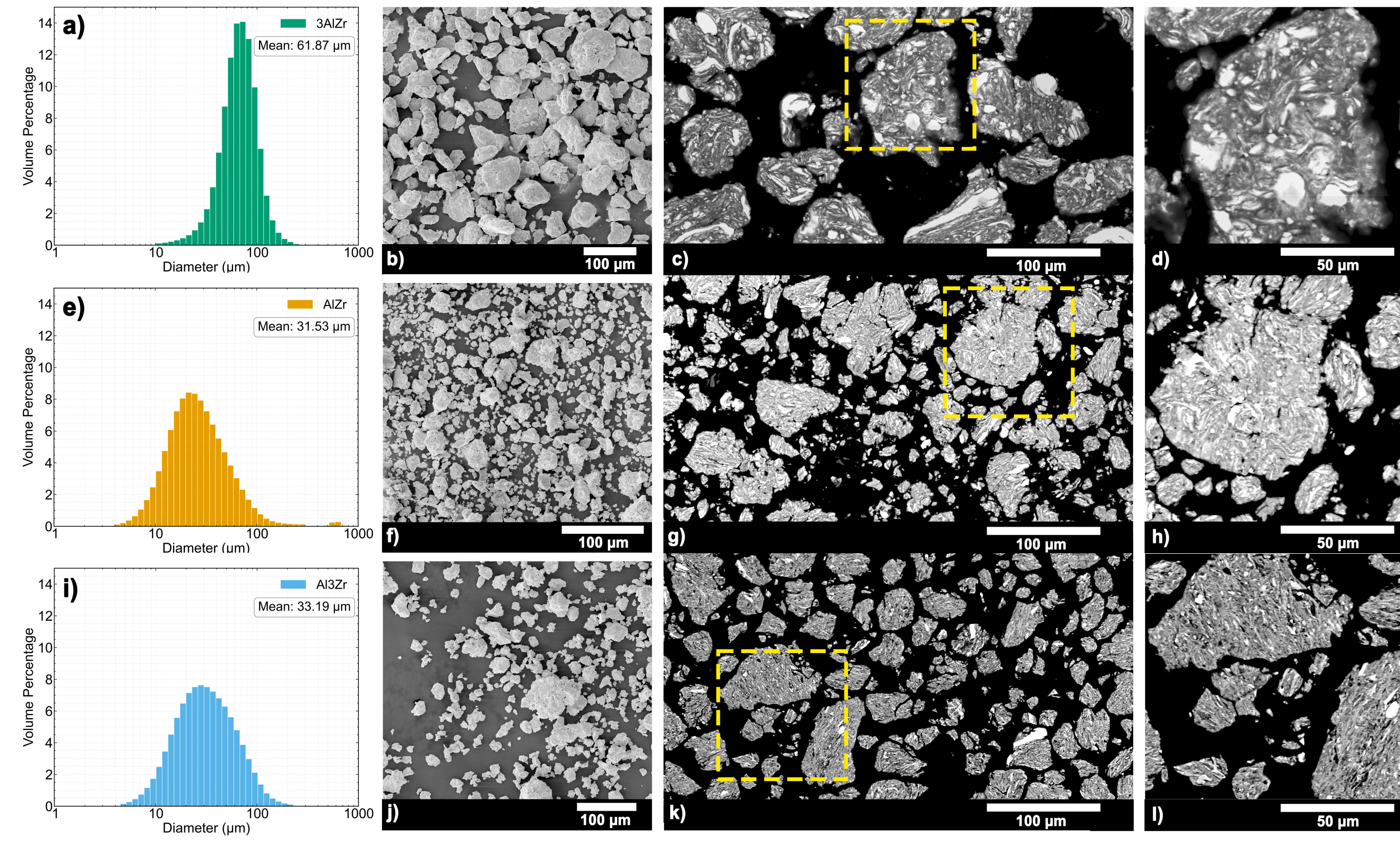}
    \caption{Particle size distribution plots for the as-milled powders: 3Al:Zr (a), Al:Zr(e), and Al:3Zr (i). Secondary electron SEM images of loose powders and a select region in the stitched backscattered electron SEM images of cross-sections for 3Al:Zr (b) to (d), Al:Zr (f) to (h), and Al:3Zr (j) to (l) showing light gray Zr inclusions embedded within a darker colored Al matrix. A selected smaller region in the select region of the stitched back-scatter images was highlighted to focus on the inclusion morphology and distribution within a sample particle: 3Al:Zr (d), Al:Zr (h), and Al:3Zr(l). The full stitched images can be seen in Supplementary Information S1.5.}
    \label{fig:Particle_Distribution}
\end{figure*}

Particle size distributions for the as-milled and sieved powders are unimodal, as shown in \cref{fig:Particle_Distribution}(a-c). The 3Al:Zr sample has a significantly larger mean particle diameter of 61.87 {\textmu}m compared to the Al:Zr and Al:3Zr powders that have similar mean particle sizes of 30.37 {\textmu}m and 31.53 {\textmu}m, respectively. The particles have rough and irregular surface morphologies but are approximately equiaxed (\cref{fig:Particle_Distribution}(d-f)). Backscattered electron (BSE) cross-sectional SEM images (\cref{fig:Particle_Distribution}(g-i)) of the powders reveal light grey Zr inclusions distributed within a darker gray Al matrix. The dark background is the mounting epoxy. In general, the Zr inclusions appear elongated and randomly distributed throughout the powders and are larger in 3Al:Zr powders compared to the Al:Zr and Al:3Zr particles. It can be inferred qualitatively from the cross-sectional BSE images that the Al:Zr and Al:3Zr powders have finer inclusions with greater interfacial area compared to the 3Al:Zr powders. 

\begin{figure*}[b!]
    \centering
    \includegraphics[width=0.8\linewidth]{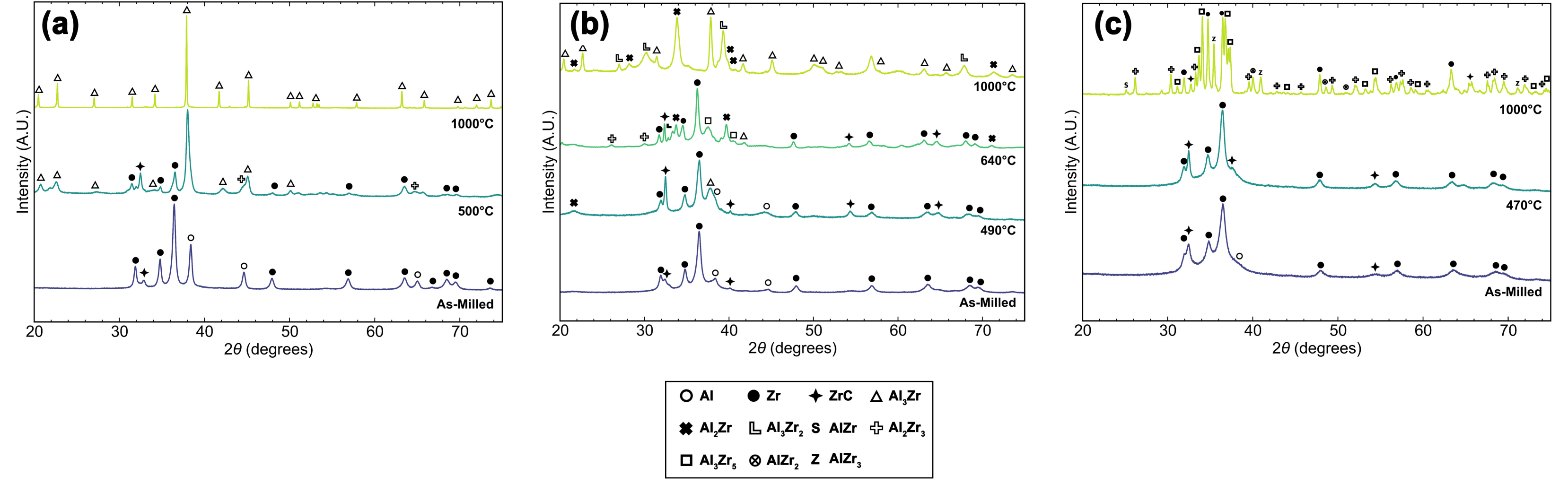}
    \caption{X-ray diffraction patterns for as-milled and annealed samples for (a) 3Al:Zr, (b) Al:Zr, and (c) Al:3Zr. For each chemistry, samples are annealed to an intermediate temperature and to 1000 °C.}
    \label{XRD_panel}
\end{figure*}

\begin{figure}[h]    
    \centering
    \includegraphics[width=1\linewidth]{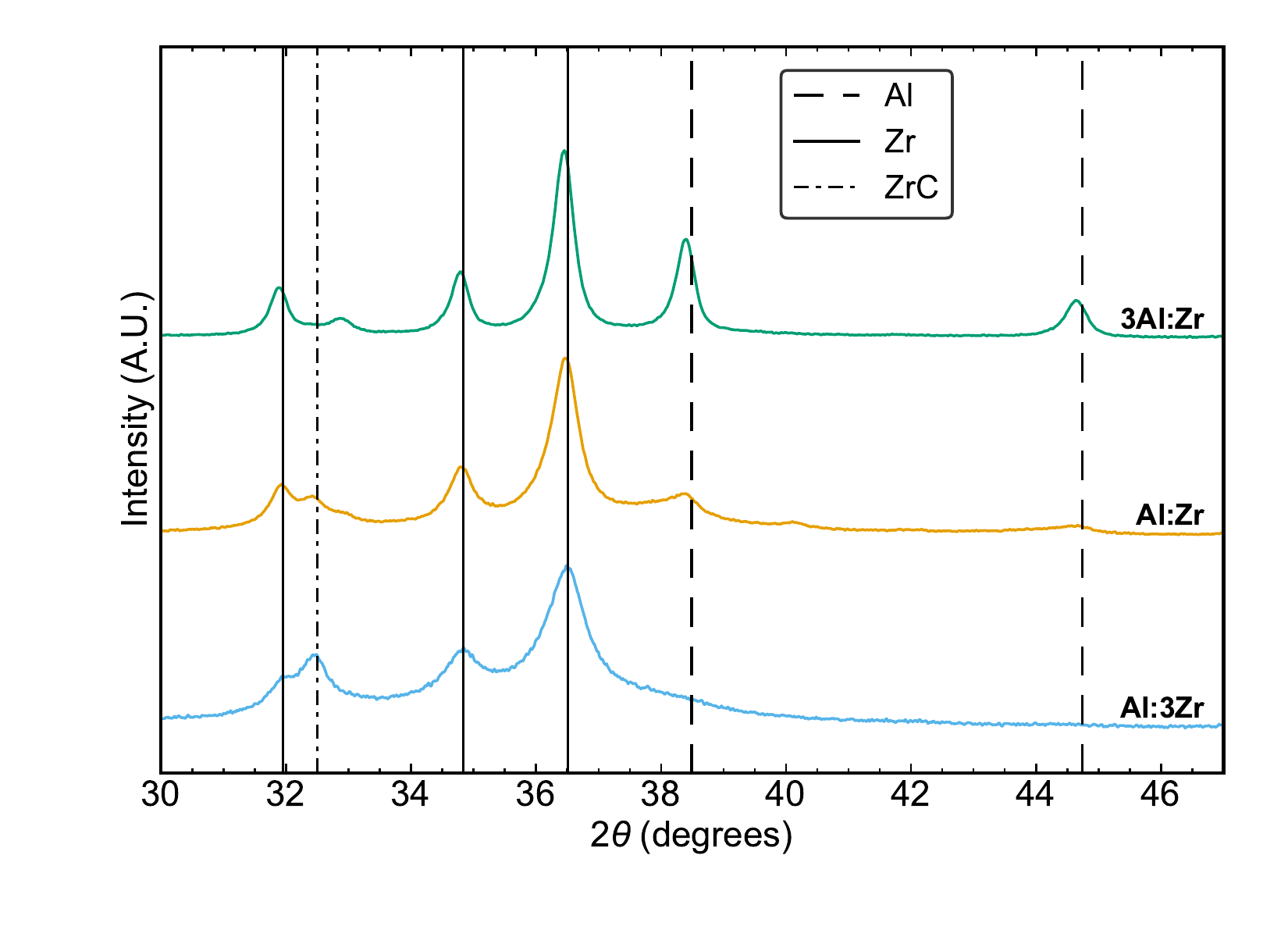}
    \caption{X-ray diffraction scans for the three as-milled powders from 30-47 degrees 2$\theta$ to show the Al and Zr peak shifts and the presence of ZrC. A different stoichiometric carbide is forming for 3Al:Zr, which is evident from the plot.}
    \label{XRD_zoom}
\end{figure}

X-ray diffraction of the as-milled and annealed samples is shown in \cref{XRD_panel}(a-c). The XRD of the as-milled powders shows the presence of elemental Al and Zr, as well as ZrC peaks, which is possibly attributed to the reaction of Zr with the hexane processing agent during milling. Predictably, the volume fraction of Al decreases in the more Zr-rich samples as seen by the shift in relative peak intensities in \cref{XRD_zoom}. 

\cref{XRD_zoom} shows that the Al (111) peaks shift towards smaller 2$\theta$  values for as-milled 3Al:Zr and Al:Zr powders, indicating intermixing of Zr into the Al matrix, which has been seen in earlier studies \cite{lakshman2019milling, polk2023comppca} of Al/Zr ball-milled powders. A peak shift may also be present in the Al:3Zr powders; however, the (111) Al reflection is too weak to confirm this with certainty. Zr peaks also shift towards lower 2$\theta$ values for the 3Al:Zr and Al:Zr powders, suggesting the possibility of the presence of interstitial carbon in the Zr lattice, as reported earlier by Polk et al. \cite{polk2023comppca}. The shift is not noticeable for the Al:3Zr powders, potentially due to the increased formation of ZrC, which reduces the amount of interstitial carbon.

As expected, intermetallic phases form in all powders as they are heated to higher temperatures. The 3Al:Zr sample exhibits evidence of its stoichiometric, stable intermetallic phase, \ce{Al3Zr}, after heating to 500 °C (49\%), and by 1000 °C, the powders have fully transformed to \ce{Al3Zr}. \cref{XRD_panel}(b) shows the Al:Zr sample has elemental Zr, ZrC, and some Al-rich intermetallic phases such as \ce{Al3Zr}, and \ce{Al2Zr} after heating to 490 °C (50\%). By 1000 °C, mostly Al-rich intermetallics are present including \ce{Al3Zr}, \ce{Al2Zr}, \ce{Al3Zr2}. Additionally, no ZrC peaks are visible, yet there is still excess Zr. Al:3Zr trends are similar to Al:Zr. After heating to 470 °C (36\%), there are only Zr and ZrC peaks present, but by 1000 °C, there are multiple Zr-rich intermetallic phases (\ce{Al2Zr3}, \ce{Al3Zr5}, \ce{AlZr2}, \ce{AlZr3}), and excess Zr. \cref{tab:XRD_phases_table} summarizes the different phases identified by XRD.

\begin{table*}[b!]
\centering
\begin{tabular}{|>{\centering\arraybackslash}m{0.1\linewidth}|>{\centering\arraybackslash}m{0.15\linewidth}|>{\centering\arraybackslash}m{0.35\linewidth}|>{\centering\arraybackslash}m{0.31\linewidth}|}
\hline
\textbf{Powder} & \textbf{As-Milled} & \textbf{Annealed at Intermediate Temperature, \textit{T}*} & \textbf{Annealed at 1000\textdegree C} \\
\hline
3Al:Zr & Al, Zr, ZrC & \ce{Al3Zr}, \ce{Al2Zr3}, Zr, ZrC & \ce{Al3Zr} \\
\hline
Al:Zr & Al, Zr, ZrC & \ce{Al3Zr}, \ce{Al2Zr}, \ce{Al3Zr5}, Zr, ZrC & \ce{Al3Zr}, \ce{Al2Zr}, \ce{Al3Zr2} \\
\hline
Al:3Zr & Al, Zr, ZrC & Al (amorphous), Zr, ZrC & \ce{AlZr3}, \ce{Al2Zr3}, \ce{Al3Zr5}, \ce{AlZr}, \ce{AlZr2}, Zr, ZrC \\
\hline
\end{tabular}
\caption{Phase composition of powders under different thermal conditions as identified by XRD. Here, \textbf{\textit{T}*} refers to the annealing temperature for the intermediate sample of each chemistry. For 3Al:Zr, 500 °C; for Al:Zr, 490 °C; and for Al:3Zr, 470 °C.}
\label{tab:XRD_phases_table}
\end{table*}

%----------------------------------------------------------------------------------------------------------------------------------------
\subsection{Thermal Analysis}
\label{subsec:thermalresults}

%\subsubsection{Intermetallic Heat of Formation} 
%\label{subsubsec:im_heat_results}

\begin{figure*}[ht]
    \centering
    \includegraphics[width=0.89\linewidth]{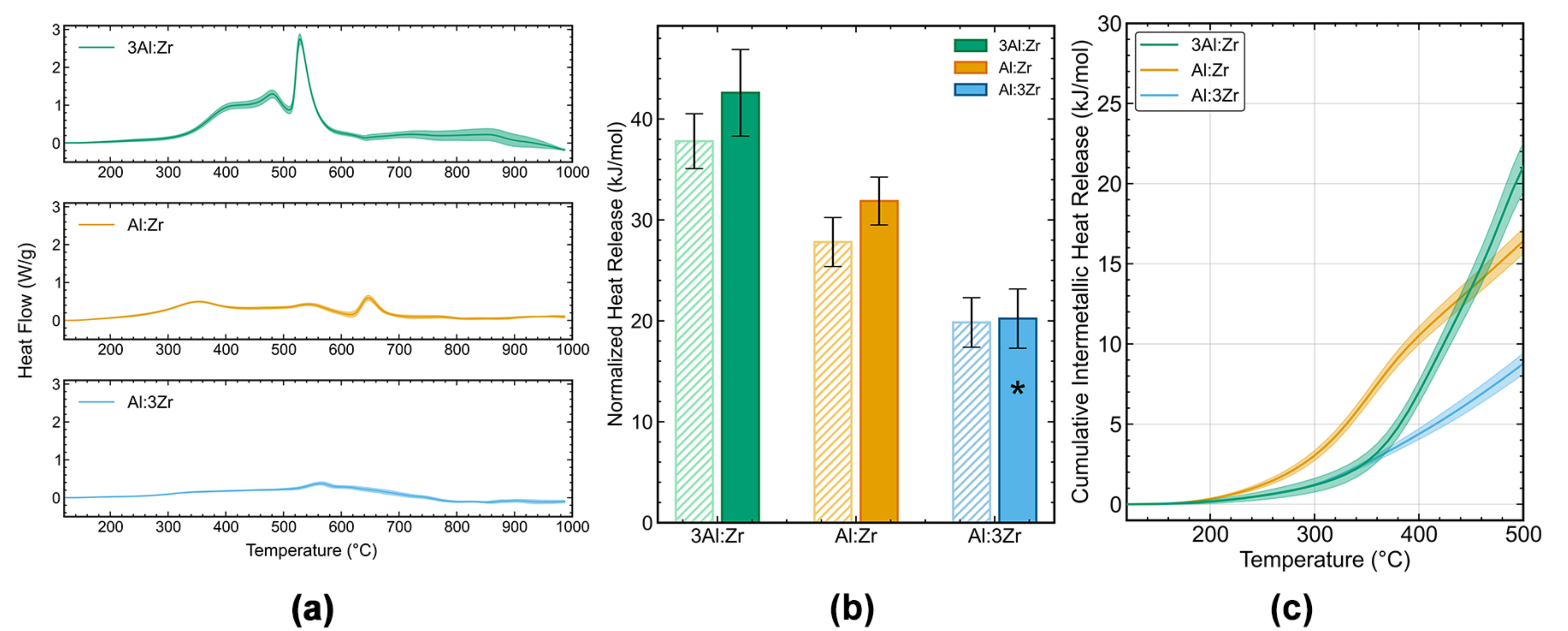}
    \caption{Results from DTA scans of as-milled powders in Ar. (a) Shaded error plots of DTA traces for each chemistry with baseline scans subtracted. (b) The normalized heat release after integrating from 150 to 700 °C (hatched bars) and from 150 to 1000 °C (solid bars). The star indicates that the integration from 150 to 1000 °C for the Al:3Zr powders stopped at $\sim$750 °C to avoid the gradual endotherm due to the transformation of $\alpha$-Zr to $\beta$-Zr as indicated by the equilibrium phase diagram \cite{murray1992alzrphaseD}. (c) Cumulative intermetallic heat release at low temperatures for each chemistry. For (a) and (c) the shaded error represents one standard deviation from the mean. For (b) the error bars represent one standard deviation from the mean.}
    \label{intermetallic_heat_figures}
\end{figure*}

\cref{intermetallic_heat_figures}(a) plots the normalized heat flow with respect to temperature for all three samples and shows exothermic peaks corresponding to Al-Zr intermixing and intermetallic formation. Relative to the normalized heat flow for 3Al:Zr, the features in the scans for Al:Zr and Al:3Zr are quite small. However, all three powders begin to release some heat by 150 °C as shown in \cref{intermetallic_heat_figures}(b), presumably due to Zr intermixing into Al at these low temperatures. Al:Zr initially releases heat faster than 3Al:Zr and Al:3Zr powders, which we attribute to finer Zr inclusions and its high theoretical heat of formation \cite{lakshman2019milling, lakshman2021evaluating, wainwright2023comparing}. 

While all three chemistries exhibit distinct exotherms, a broad endotherm begins at 750 °C for the Al:3Zr powders and is present until the end of the measurement at 1000 °C. The endotherm is attributed to intermetallics breaking down and forming more $\alpha$-Zr in the individual Al:3Zr powders that are Zr-rich relative to the average 1:3 composition. As predicted by the equilibrium Al-Zr phase diagram \cite{murray1992alzrphaseD}, the $\alpha$-Zr transforms to $\beta$-Zr on heating above 910 °C and continues to grow in volume fraction in these individual Zr-rich powders. When calculating the total intermetallic heats released up to 700 °C and 1000 °C, this endotherm is ignored and the calculated values are plotted in \cref{intermetallic_heat_figures}(b). A heat of 42.6 ± 4.3 kJ/mol is calculated for 3Al:Zr powders scanned to 1000 °C, which approaches the reported heat of formation of 53 ± 2 kJ/mol for 3Al:Zr \cite{fisher2013phase, macikag2018enthalpy}. The calculated heats decrease with higher Zr concentrations due in part to less complete reactions. 

\begin{figure*}[h!]
    \centering    
    \includegraphics[width=0.89\linewidth]{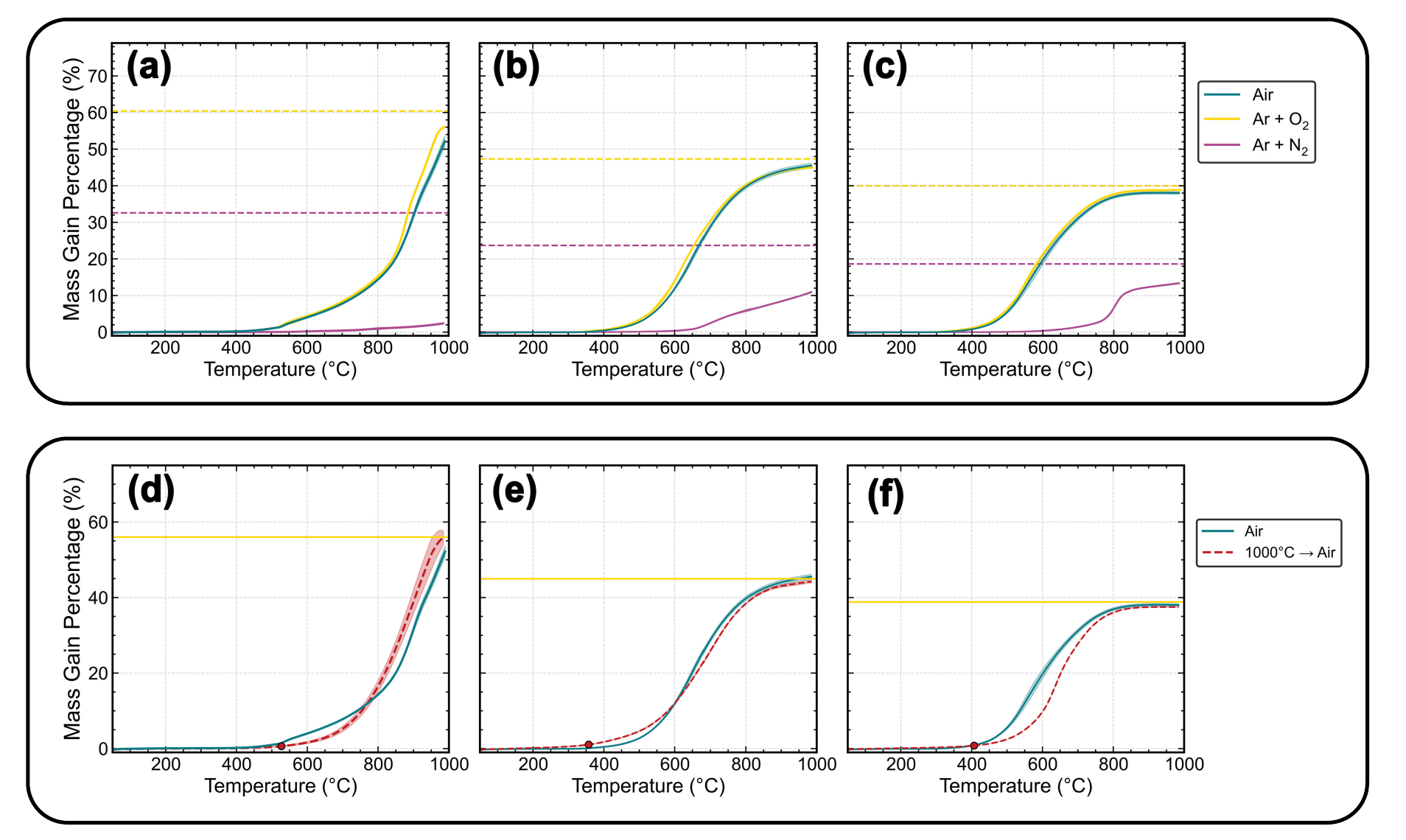}
    \caption{The mass gain percentage versus temperature for as-milled powders run in Ar + \ce{O2} and Ar + \ce{N2} for (a) 3Al:Zr (b) Al:Zr and (c) Al:3Zr powders. The dashed horizontal lines in plots (a-c) represent the total theoretical mass gain from oxidation (gold) or nitridation (purple), respectively. The mass gain percentage versus temperature for powders scanned in air, both as-milled powders and those annealed (1000 °C) in Ar for (d) 3Al:Zr (e) Al:Zr and (f) Al:3Zr powders. The solid gold lines in plots d-f denote the mean mass gain at 1000 °C for the as-milled powders heated in Ar + \ce{O2}, and the red circles indicate the onset temperature reported by Paljevic \cite{paljevic1994high} for the most easily oxidized phase found in XRD scans of each sample heated in Ar.}
    \label{thermal_analysis_mass_gain}
\end{figure*}

%\subsubsection{Heats of Oxidation and Nitridation} 
%\label{subsubsec:ox_nit_results}

The mass gain percentage in oxidizing (Ar + \ce{O2}) and nitriding (Ar + \ce{N2}) environments is plotted in  \cref{thermal_analysis_mass_gain}(a-c). The onset of oxidation occurs earlier as the Zr content increases with onset temperatures of 402 °C, 331 °C, and 305 °C for 3Al:Zr, Al:Zr, and Al:3Zr samples, respectively. Similarly, the Zr-rich Al:3Zr powders appear to plateau in mass gain around 880 °C, while the 3Al:Zr and Al:Zr powders continue to gain mass and are assumed to still be oxidizing at 1000 °C. However, all samples oxidize to within 4\% of their theoretical mass gains that are shown by dashed lines for both stoichiometric oxides and nitrides \cite{xin2020phase} in \cref{thermal_analysis_mass_gain}(a-c). We also observe a steep increase in the rate of oxidation for 3Al:Zr powders around 850 °C. 

In contrast to samples heated in Ar + \ce{O2}, samples heated in Ar + \ce{N2} exhibit significantly lower mass gain percentages, with observed values falling 5–30\% below theoretical expectations. The onset of nitridation does not decrease with increasing Zr content as seen for oxidation.  Instead, it is earliest for Al:Zr, which begins nitridation around 380 °C and only starts for 3Al:Zr powders at 478 °C and for Al:3Zr powders at 432 °C. However, the overall percentage mass gain does increase consistently with Zr content and is greatest at 1000 °C for Al:3Zr powders with the highest value of 13.6\%. This is due in part to a significant rise in the rate of mass gain around 800 °C for the Zr-rich powders. In general, the rates of oxidation and nitridation appear to increase after Al melts. 

%Mass Gain in Air

%\subsubsection{DTA/TGA In Air and Ar-Air} 
%\label{subsubsec:air_results}

Next, we compare DTA/TGA data in \cref{thermal_analysis_mass_gain}(d-f) for as-milled powders and powders scanned to 1000 °C in Ar. These experiments were conducted to investigate the impact of preformed intermetallic phases on mass gain when scanning to 1000 °C in air. In general, the annealed powders have a similar total mass gain compared to the as-milled powders, particularly when focusing on the mass gain at 1000 °C. The mass gain in air at 1000 °C also matches the mass gain in Ar + \ce{O2} as shown by the yellow horizontal lines. However, the onset of mass gain is delayed by $\sim$100 °C in annealed Al-rich and Zr-rich powders compared to the as-milled powders, suggesting the presence of specific intermetallics in those powders delays the start of oxidation in 3Al:Zr and Al:3Zr samples. In contrast, the annealed Al:Zr powders start gaining mass sooner than the as-milled Al:Zr powders.  

%----------------------------------------------------------------------------------------------------------------------------------------

\subsection{Wire Ignition Results}
\label{subsec:ignitionresults}

\begin{figure}[h!]
    \centering
    \includegraphics[width=1\linewidth]{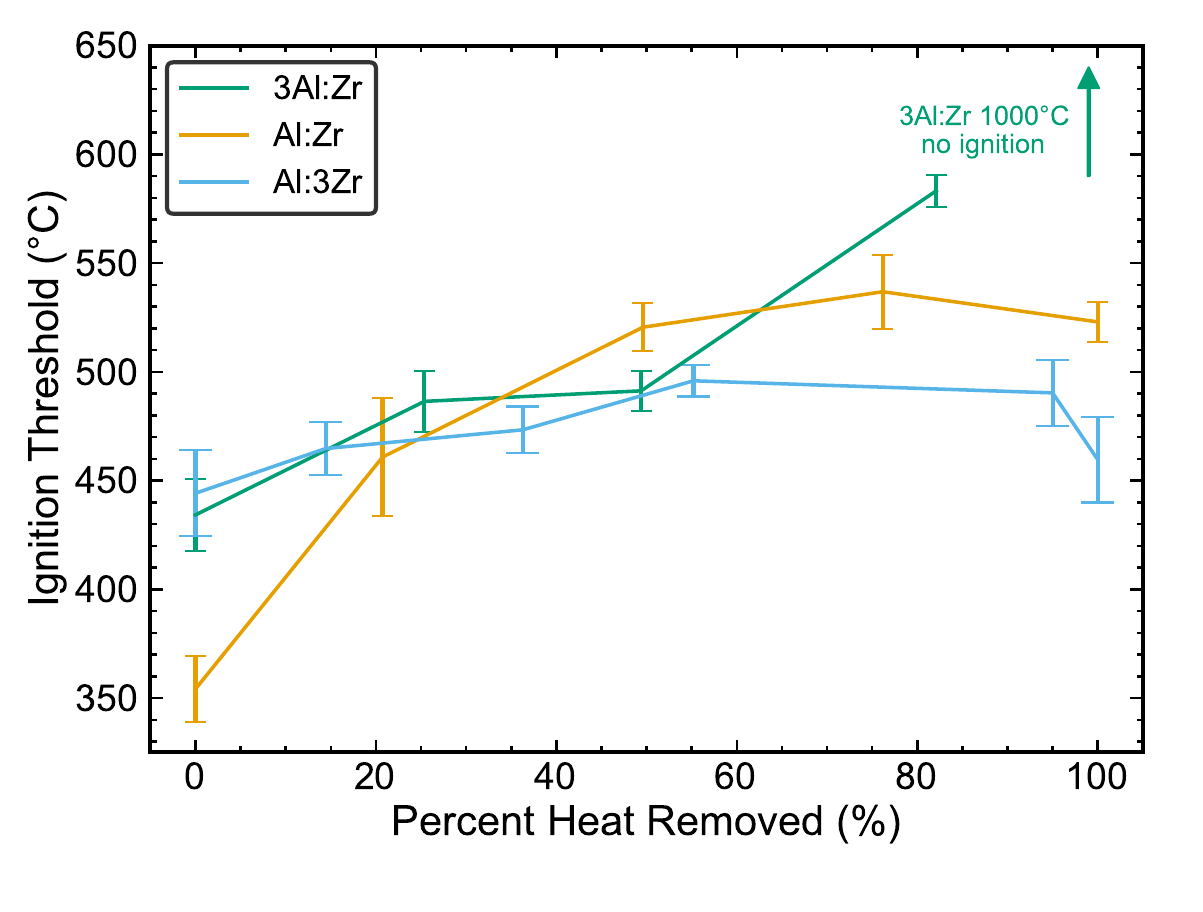}
    \caption{Wire ignition results in air for all as-milled and annealed samples. The 3Al:Zr 1000 °C sample was unable to ignite before the nichrome filament broke. Error bars represent one standard deviation from the mean.}
    \label{fig:wire_ignition_air}
\end{figure}

\begin{figure*}[h!]
    \centering
    \includegraphics[width=0.89\linewidth]{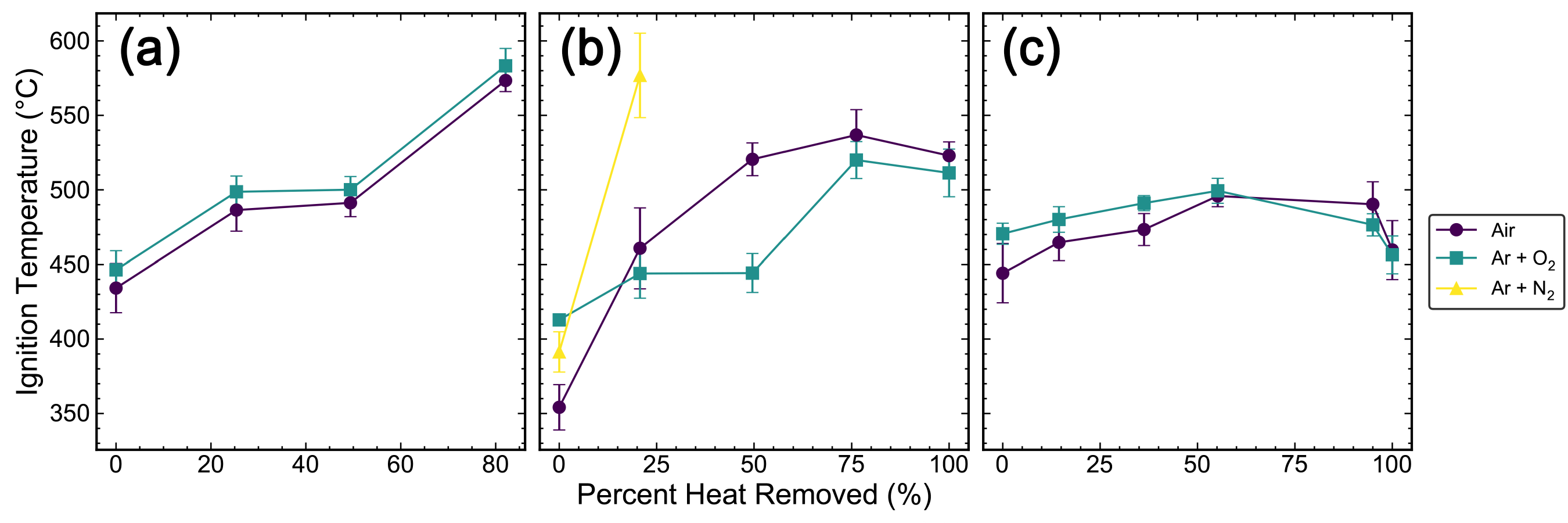}
    \caption{Comparison of wire ignition thresholds in different gas environments for as-milled and annealed (a) 3Al:Zr (b) Al:Zr and (c) Al:3Zr powders. 3Al:Zr 1000 °C did not ignite in air or Ar + \ce{O2}. None of the samples for 3Al:Zr or Al:3Zr ignited in Ar + \ce{N2}. Error bars represent one standard deviation from the mean. Note: For a few samples, the error bars are smaller than the marker, which prevents them from being visible.}
    \label{fig:wire_ignition_all_envs}
\end{figure*}

The ignition temperatures and standard deviations are plotted in \cref{fig:wire_ignition_air} for as-milled and annealed powders ignited in air. The as-milled Al:Zr powders have a significantly lower ignition threshold (354.2 ± 15.22  °C) compared to the as-milled 3Al:Zr (434.2 ± 16.53  °C) and Al:3Zr powders (444.2 ± 19.83  °C), which are similar. Annealing in Ar and the formation of intermetallics increases ignition thresholds in general. Ignition temperatures rise the most for Al:Zr powders as the first 50 percent of heat is removed, but then rise significantly more for the 3Al:Zr powders as the last 50 percent of heat is removed. 

Above 50\% heat removal, the 3Al:Zr powders can no longer ignite prior to the 28-gauge nichrome filament breaking at approximately 900 °C. Thus, the ignition threshold for the annealed 3Al:Zr powders and the \ce{Al3Zr} intermetallic phase that forms is likely above 900 °C.  In sharp contrast to the Al-rich powders, the ignition temperatures of the Al:Zr and Al:3Zr powders plateau and then fall as the last 50 percent of heat is removed. Ignition temperatures for Al:Zr powders level off around 520-540 °C while they reach a maximum of 496 °C for Al:3Zr samples before decreasing after annealing to 1000 °C. 

Wire ignition tests performed in air, \ce{Ar} + \ce{O2}, and \ce{Ar} + \ce{N2} on as-milled and annealed powders are plotted in \cref{fig:wire_ignition_all_envs} for each chemistry. The ignition thresholds in air and \ce{Ar} + \ce{O2} are similar for 3Al:Zr and Al:3Zr powders, differing by less than 25 °C for all samples. However, they differ by as much as 70 °C for the Al:Zr powders depending on the degree of annealing. The as-milled Al:Zr powders have 50 °C lower ignition temperatures in air compared to \ce{Ar} + \ce{O2}, but have up to 70 °C higher ignition temperatures in air compared to \ce{Ar} + \ce{O2} once 50\% of the intermetallic heat is removed by annealing. However, once the Al:Zr powders are scanned to 1000 °C in Ar for 100\% removal,  their ignition temperatures are similar in air and Ar + \ce{O2}. 

Differences in ignition thresholds are more dramatic in \ce{Ar} + \ce{N2} where the as-milled 3Al:Zr and Al:3Zr powders glowed dimly, but did not ignite, indicating these two chemistries require some degree of oxidation to ignite. In sharp contrast, the as-milled Al:Zr powders ignite in \ce{Ar} + \ce{N2} at similar or lower temperatures than \ce{Ar} + \ce{O2}. However, once annealing occurs, the ignition temperatures for the Al:Zr powders rise sharply in \ce{Ar} + \ce{N2} and cannot be measured above 25 percent heat removal, suggesting ignition temperatures are above 900 °C.

%----------------------------------------------------------------------------------------------------------------------------------------
\subsection{Combustion Results}

\begin{figure*}[b!]
    \centering
    \includegraphics[width=0.9\linewidth]{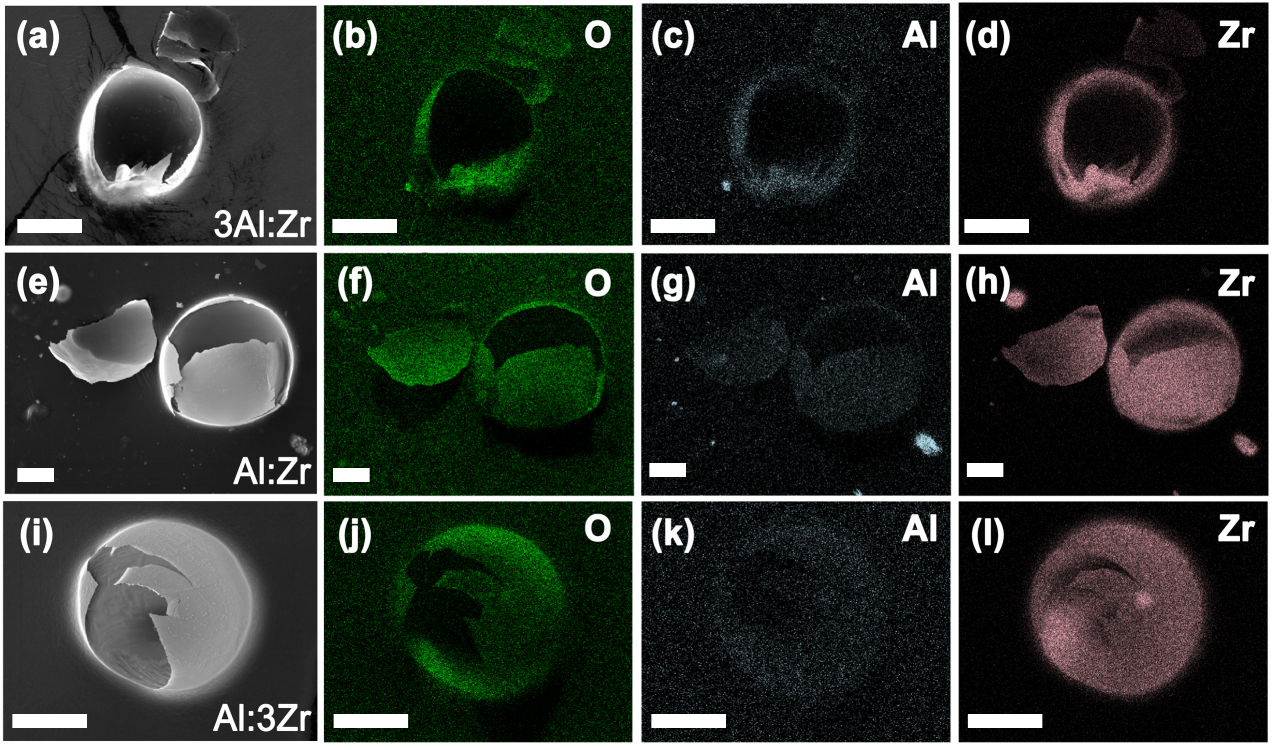}
    \caption{EDS of combustion products captured six inches above the plasma ignition source for 3Al:Zr (a-d), Al:Zr (e-h), and Al:3Zr (i-l) powders. Images in the first column (a,e,i) show a secondary electron SEM image of each product, the second column (b,f,j) shows the oxygen content in each product, while (c,g,k) shows the Al content, and (d,h,l) shows the Zr content from an EDS map. There are some unignited particles in upper left and lower right corners for the Al:Zr images (e-h). Scale bars represent 20 microns for each image.}
    \label{eds_figure}
\end{figure*}

% Qualitative observations? 
SEM imaging and Energy Dispersive Spectroscopy (EDS) were performed on combustion products of as-milled powders that were collected six inches above the plasma in which they were ignited. The collected combustion products all showed evidence of microexploded fragments and/or hollow zirconia spheres as presented in \cref{eds_figure}. This supports the earlier report by Wainwright et al. \cite{wainwright2018gasenvs} that Al:Zr ball-milled powders combust in a dual-phase process, whereby Al evaporates and oxidizes in the vapor phase while there is nitridation of the condensed Al-Zr liquid solution. Zr rapidly oxidizes in the condensed state after Al has finished evaporating, leading to nucleation and growth of \ce{N2} bubbles that form hollow oxide spheres if they grow slowly and microexplosions if they grow rapidly \cite{wainwright2018gasenvs}.

% Burn Duration
%\subsubsection{Burn Duration} 
%\label{subsubsec:burn_duration}

\begin{figure*}[b!]
    \centering
    \includegraphics[width=0.8\linewidth]{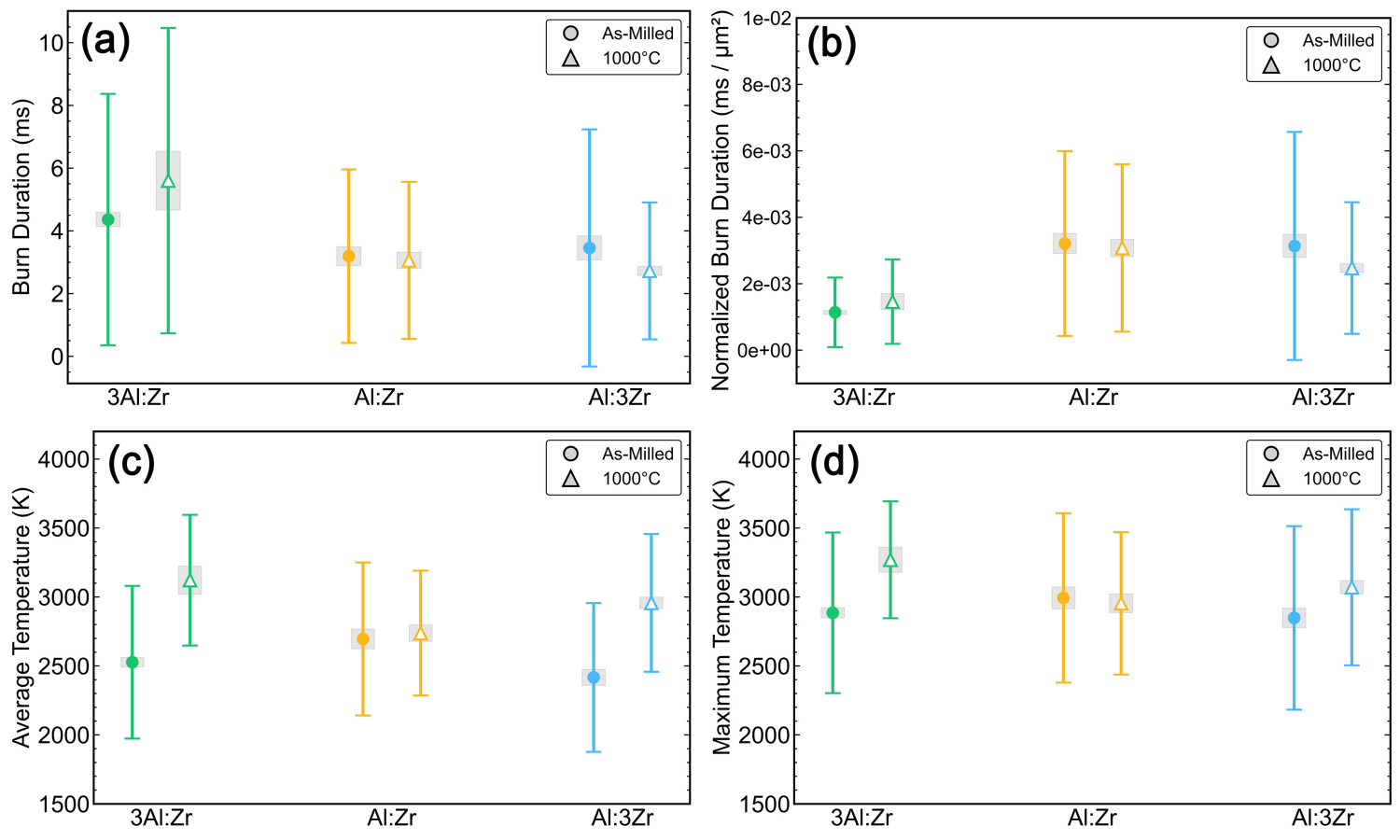}
    \caption{The (a) non-normalized and (b) normalized average burn durations for as-milled powders and ones scanned to 1000 °C for each chemistry. Plot (b) is normalized by the mean particle diameter squared based on the particle size analysis results. Plots (c) and (d) show the average (c) and average maximum (d) temperatures for the combusted particles for each chemistry and condition. For all plots, error bars represent one standard deviation, and the shaded boxes represent the bounds of the 95\% confidence interval of the mean values.}
    \label{fig:combustion_results}
\end{figure*}

The burn durations of as-milled powders and those scanned to 1000 °C in Ar typically range from 2 to 10 ms and are plotted in \cref{fig:combustion_results}(a). In compiling these data sets, particles with less than 1 ms of data were discarded, as many of these data points resulted from tracking errors. Moreover, particles that burned for more than 30 ms were treated as outliers and were discarded. This filtering to less than 30 ms accounts for the possibility that some particles sintered together and formed larger particles with extended burn durations. \cref{tab:combustion_summary} reports the number of particles that passed filtering and were considered for analysis for each chemistry and condition tested. See Supplementary Information 1.4.2 for more details regarding the filtering criteria used for data analysis.

In general, annealing as-milled powders to 1000 °C produced different degrees of change in burn duration for each chemistry (\cref{fig:combustion_results}(a)). The 3Al:Zr powders scanned to 1000 °C in Ar show the most variability in burn duration, but this could be due to the fact that they are difficult to ignite and very few annealed 3Al:Zr particles combusted and passed filtering (less than one hundred) compared to the other powders which all had hundreds of data points. We observed that the as-milled and annealed 3Al:Zr samples burn longer than the Al:Zr and Al:3Zr powders; however, part of this difference can be attributed to their larger average size. Literature suggests that burn time (\(\tau_{\text{burn}}\)) for microscale powders is related to the particle diameter (\(D\)) by the following power-law relationship: \(\tau_{\text{burn}} \sim a D^n\), where \(a\) is a proportionality constant and \(n\) is the scaling exponent \cite{lynch2009correlation, wilson1971experimental}. 

To account for this dependence, durations were normalized by the square of the mean, as-milled particle diameter \cite{wilson1971experimental, shoshin2006particle} and are shown in \cref{fig:combustion_results}(b). The choice of normalizing by mean diameter squared is based on the estimate predicted from diffusion-limited droplet theory for larger micron-scale Al particles \cite{lynch2009correlation}. Experiments of Al particles have shown that this exponent is lower than expected \cite{beckstead2005correlating}, and will change for smaller particles \cite{gill2010combustion}. There would also be a deviation from this value for particles that burned in a condensed state, like those with more Zr. The mean starting diameters were used because the particles in the imaging channel are frequently saturated, making their in-situ size determination inaccurate. However, using the initial mean particle diameter may also be misleading if certain particle sizes ignite and combust more easily than others (i.e., smaller particles). This would be particularly true for ball-milled powders that produce a broad range of powder sizes like the ones used for this study. 

Examining the normalized results in \cref{fig:combustion_results}(b), we find that the as-milled and annealed 3Al:Zr particles burn for shorter durations compared to the Al:Zr and Al:3Zr powders, opposite to the non-normalized results in \cref{fig:combustion_results}(a). Looking at the effect of annealing on burn duration, we found that annealing substantially increased the burn duration for 3Al:Zr and decreased it for Al:3Zr powders, but had a minimal effect on burn durations for Al:Zr powders. 

\begin{table*}[h!]
\centering
\resizebox{1\linewidth}{!}{%
\begin{tabular}{|l|c|c|c|c|c|c|}
\hline
\textbf{Powder} & 
\textbf{\begin{tabular}[c]{@{}c@{}}Burn Duration\\ (ms)\end{tabular}} & 
\textbf{\begin{tabular}[c]{@{}c@{}}Normalized Burn\\ Duration (1e-3 ms/um\textsuperscript{2})\end{tabular}} & 
\textbf{\begin{tabular}[c]{@{}c@{}}Average\\ Temperature (K)\end{tabular}} & 
\textbf{\begin{tabular}[c]{@{}c@{}}Maximum\\ Temperature (K)\end{tabular}} & 
\textbf{\begin{tabular}[c]{@{}c@{}}Microexplosion \\ Percentage (\%)\end{tabular}} & 
\textbf{\begin{tabular}[c]{@{}c@{}} \# Particles \end{tabular}} \\ \hline
\textbf{3Al:Zr As-Milled} & 4.36 ± 4.00 & 1.14 ± 1.05 & 2432 ± 573 & 2885 ± 582 & 61.6 ± 2.4 & (1201, 984) \\ \hline
\textbf{3Al:Zr 1000C} & 5.60 ± 4.87 & 1.46 ± 1.27 & 3030 ± 466 & 3270 ± 424 & 46.1 ± 8.2 & (106, 86) \\ \hline
\textbf{Al:Zr As-Milled} & 3.19 ± 2.77 & 3.21 ± 2.78 & 2598 ± 580 & 2993 ± 616 & 66.3 ± 3.5 & (329, 230) \\ \hline
\textbf{Al:Zr 1000C} & 3.06 ± 2.50 & 3.08 ± 2.52 & 2642 ± 484 & 2954 ± 517 & 73.9 ± 3.6 & (362, 220) \\ \hline
\textbf{Al:3Zr As-Milled} & 3.46 ± 3.78 & 3.14 ± 3.43 & 2376 ± 525 & 2848 ± 666 & 54.8 ± 4.6 & (379, 343) \\ \hline
\textbf{Al:3Zr 1000C} & 2.73 ± 2.18 & 2.47 ± 1.98 & 2880 ± 616 & 3070 ± 566 & 66.9 ± 2.4 & (878, 543) \\ \hline
\end{tabular}%
}
\caption{Table summarizing key metrics for the different powders calculated using SHEAR data. For every metric except microexplosion percentage and number of particles, the first number is the mean value and the second is the standard deviation. The error for microexplosion percentage represents bootstrapped 95\% confidence intervals. The last column lists the number of particles that were considered for the burn duration and temperature calculations, respectively.}
\label{tab:combustion_summary}
\end{table*}

% Avg, Max Temperature
%\subsubsection{Combustion Temperatures} 
%\label{subsubsec:combustion_temperatures}

The average and maximum temperatures for the three chemistries and conditions are shown in \cref{fig:combustion_results} and \cref{tab:combustion_summary}. The calculated average temperatures shown in \cref{fig:combustion_results} and \cref{tab:combustion_summary} are particle-mean (unweighted) averages. The weighted, frame-mean average combustion temperatures are given in Supplementary Information 1.4.3, and show that using a weighted average slightly changes the average temperatures, but does not drastically affect any of the observed trends shown here. In general, across all chemistries and conditions, the particles burn at temperatures ranging from 2000 to 3500 K. The standard deviations are relatively large, varying by 450 to 650 K (about 20-25\% of the averages), and the average maximum temperatures are 100–400 K higher than the average temperature for each chemistry and processing condition. Comparing as-milled versus annealed powders, both the 3Al:Zr 1000 °C and Al:3Zr 1000 °C powders burned at substantially higher average temperatures, approximately 550–600 K higher, while the Al:Zr 1000 °C powder burned at a similar average temperature to its as-milled counterpart. Meanwhile, the difference in average maximum temperatures was less pronounced but still evident, with annealed 3Al:Zr and Al:3Zr powders burning 200–400 K higher than the as-milled 3Al:Zr and Al:3Zr powders. Analysis of the average and maximum combustion temperatures reveals that as-milled powders differ by only 150–300 K across varying Zr concentrations.

\begin{figure}[ht]
    \centering
    \includegraphics[width=1\linewidth]{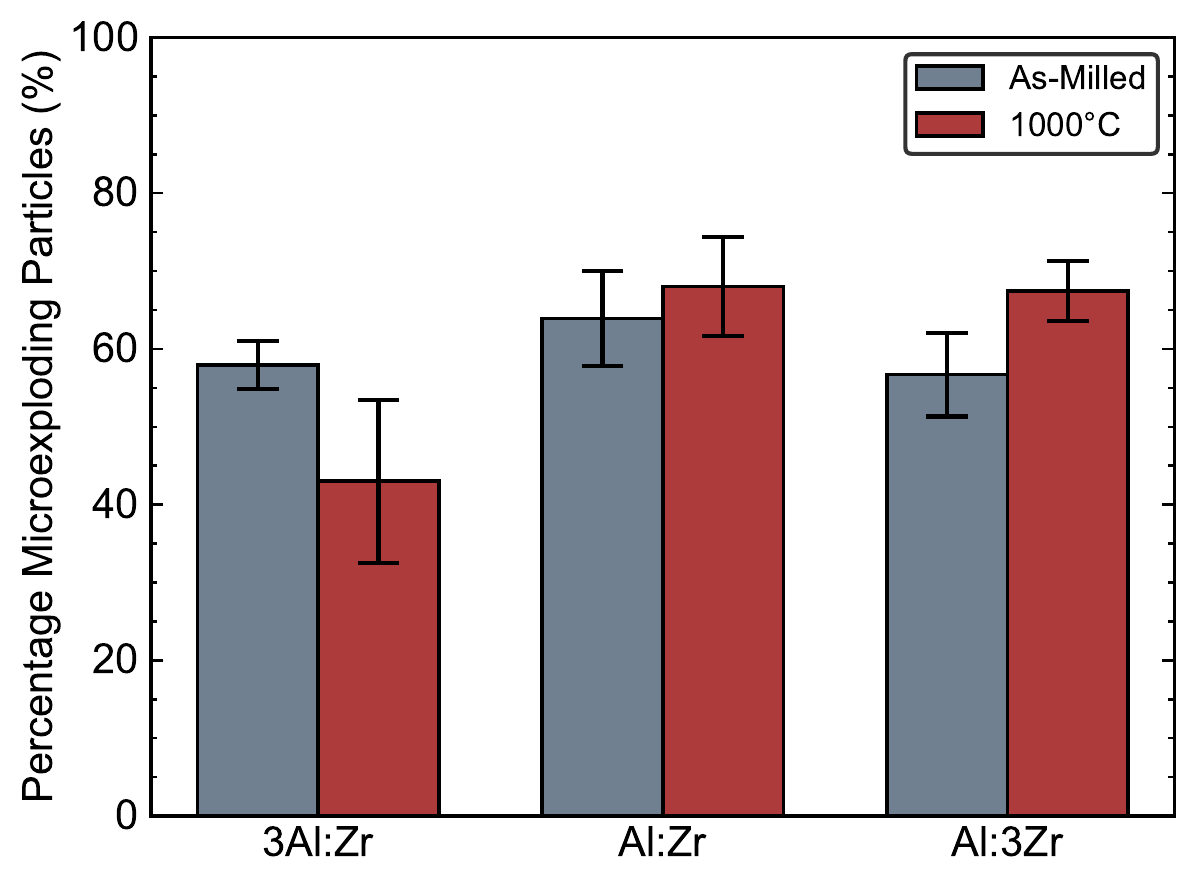}
    \caption{Microexplosion frequency for all powders tested in the combustion experiments. Error bars represent 95\% confidence intervals found by averaging results from 1000 randomly bootstrapped samples.}
    \label{fig:microexplosion_plot}
\end{figure}

Elemental and composite micron-sized metal powders often experience microexplosions during combustion, resulting in violent fragmentation of the droplet \cite{mcnanna2025disruptive}. All particles that passed the filtering criteria specified in Supplementary Information Section S1.4.2 were included in the microexplosion analysis, the results of which are shown in \cref{fig:microexplosion_plot}. 

Despite not all following the same trend, we observed a statistically significant difference in microexplosion frequency between as-milled and annealed powders for each chemistry (p-values: 0.0002, 0.0023, and 4.37$\times 10^{-6}$ for 3Al:Zr, Al:Zr, and Al:3Zr respectively), though the trend (increasing or decreasing frequency) varied based on chemistry. There was a decrease in microexplosion frequency between as-milled and annealed 3Al:Zr, but Al:Zr and Al:3Zr saw an increase in microexplosion frequency for annealed powders. There was no consistent trend in microexplosion frequency with Zr content. 3Al:Zr and Al:3Zr powders microexploded less frequently compared to Al:Zr powders. Overall, the powders across all chemistries and conditions often microexploded, a phenomenon that has been qualitatively observed in other studies. 

We also investigated whether microexplosion frequency could influence burn duration, which was observed by Zhou et al. \cite{zhou2025fundamental} for atomized Al-Li alloys with Li content ranging from 2.5 to 10 wt\%. Combining all particles that passed filtering into a single dataset, no correlation was found between microexplosion events and burn duration (Pearson's correlation coefficient: 0.0153, p-value: 0.4536). Correlation testing was performed on each specific powder as well, but still, no significant correlations were found. 

\section{Discussion}
\label{sec:Discussion}

\subsection{Ignition Properties}
\label{subsec:discussion_ignition}

\subsubsection{As-Milled Ignition Thresholds}
\label{subsec:discussion_ignition_thresholds}

The temperatures reported in \cref{fig:wire_ignition_air} and \cref{fig:wire_ignition_all_envs} represent the lowest temperatures at which exothermic reactions of the powders become self-sustaining in various environments. The ignition thresholds reported for the as-milled powders in air are in reasonable agreement with previous studies of Al/Zr ball-milled powders \cite{wainwright2020microstructure, wainwright2023comparing, lakshman2019milling,polk2023comppca} with any differences likely due to variations in the amount of stored chemical energy, the degree of premixing, or the refinement of the Zr inclusions. To validate that heat released from intermixing and intermetallic formation is the most important factor in driving ignition, we calculated equivalent ignition temperatures for the theoretical case where ignition studies are conducted at slow heating rates that match the DTA/TGA scans. We did this using the Kissinger method as reported in other studies \cite{polk2023comppca, swaminathan2013studying}. Zr intermixing into Al is thought to control the initial heat release and the activation energy for this intermixing was estimated to be 2.4 eV/atom (231.58 kJ/mol) based on studies done by Fisher et al. \cite{fisher2013phase}. 

The equivalent ignition temperatures for each chemistry at slow heating rates are given in \cref{tab:equivalent_ignition}. Next, we calculated the energy released due to intermixing and intermetallic formation compared to the energy released by oxidation and nitridation. The heats of oxidation were calculated by assuming that all mass gain up to 400 °C is due to zirconia or zirconium nitride formation based on the assumptions stated in Wainwright et al. \cite{wainwright2020microstructure}. At these low temperatures, \ce{ZrO2} should be the primary source of mass gain in \ce{Ar} + \ce{O2} and ZrN should be the primary source of mass gain in \ce{Ar} + \ce{N2} \cite{shu2020nonstoichiometric}. 

\begin{table}[h!]
\centering
\begin{tabular}{|>{\centering\arraybackslash}m{0.2\linewidth}|>{\centering\arraybackslash}m{0.2\linewidth}|>{\centering\arraybackslash}m{0.2\linewidth}|>{\centering\arraybackslash}m{0.2\linewidth}|}
\hline
\textbf{Chemistry} & \textbf{Equivalent Ignition Temperature at 0.33 °C/s ( °C)} & \textbf{Intermetallic Heat Release (kJ/mol)} & \textbf{Standard Deviation (kJ/mol)} \\ \hline
3Al:Zr & 278 & 0.86 & 0.44 \\ \hline
Al:Zr & 228 & 0.74 & 0.13 \\ \hline
Al:3Zr & 285 & 0.95 & 0.04 \\ \hline
\end{tabular}
\caption{Results from Kissinger Analysis performed using an activation energy of $E_a$ = 231.58 kJ/mol from Fisher \cite{fisher2013phase} and the heating rates from the wire ignition tests performed in air for as-milled powders.}
\label{tab:equivalent_ignition}
\end{table}

\begin{figure*}[ht]
    \centering
    \includegraphics[width=1\linewidth]{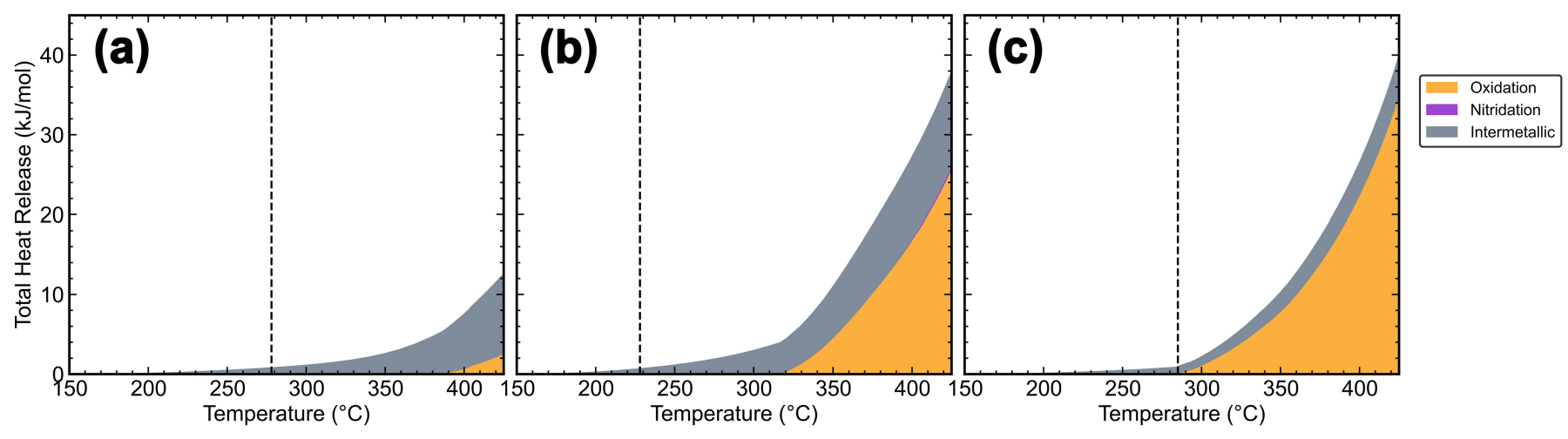}
    \caption{The heats of intermetallic heat release, oxidation, and nitridation were calculated using data from the DTA/TGA scans in different gas environments as a function of temperature. Black dotted lines represent the equivalent ignition temperature found using a Kissinger analysis.}
    \label{fig:thermal_analysis_summary_fig}
\end{figure*}

Given that the heat of formation for \ce{ZrO2} is 1100.4 kJ/mol \cite{chase1998nist}, we found that for every 1 gram of mass gain, there will be 34.3 kJ of exothermic heat released. Similarly, the heat released due to ZrN formation was found to be 14.79 kJ per 1 gram of mass gain in \ce{Ar} + \ce{N2} \cite{chase1998nist}. The results from these calculations are shown in \cref{fig:thermal_analysis_summary_fig} with dashed lines to denote the equivalent ignition threshold estimated for each chemistry. Note that the heat of nitridation is, in fact, plotted in \cref{thermal_analysis_mass_gain}(g-i); however, the contribution is so small at these low temperatures that it is difficult to see. \cref{fig:thermal_analysis_summary_fig} clearly shows that ignition occurs prior to any oxidation or nitridation for each chemistry, thus validating that the heat released from intermixing and intermetallic formation is driving ignition for as-milled powders in air. 

Moreover, we also calculated the heat from intermixing and intermetallic formation up until the equivalent ignition temperature, which is given in \cref{tab:equivalent_ignition}. We found that less than 1 kJ/mol of energy release is needed to ignite all three chemistries. Thus, in air, the onset of ignition appears to be dominated by the initial intermixing of Zr into Al and the corresponding heat release as suggested in previous studies \cite{wainwright2020microstructure, lakshman2019milling, wainwright2023comparing}. In support of this argument, we note that the Al:Zr powders, which exhibit the earliest heat release due to intermixing, likely resulting from refined Zr inclusions, also have the lowest ignition thresholds. 

% All as-milled powders ignite at lower temperatures in air than ArO2 or ArN2
While intermixing appears to stimulate ignition, the atmosphere also plays a role in determining ignition thresholds, as shown in \cref{fig:wire_ignition_all_envs}, particularly for the Al:Zr powders. For all chemistries, the ignition threshold in air is lower than in \ce{Ar} + \ce{O2}, suggesting nitrogen helps accelerate and extend the initial exothermic reactions into full combustion (oxidation and nitridation). The difference between air and \ce{Ar} + \ce{O2} is most pronounced for the Al:Zr sample, which also shows the earliest onset of mass gain during TGA scans in \ce{Ar} + \ce{N2}. However, since mild combustion only occurs for the as-milled and 25 percent Al:Zr powders in \ce{Ar} + \ce{N2}, oxygen is critical for enabling complete combustion. Supplementary Information Section 1.6 presents wire ignition still images comparing the as-milled Al:Zr powder in air, \ce{Ar} + \ce{O2}, and \ce{Ar} + \ce{N2} atmospheres, highlighting the visibly milder ignition behavior in the non-oxygenated environment relative to the oxygenated environments. Both 3Al:Zr and Al:3Zr did not ignite in \ce{Ar} + \ce{N2}, and correspondingly demonstrate limited nitridation at low temperatures in TGA experiments.

\subsubsection{Annealed Ignition Thresholds}
\label{subsec:discussion_as_milled_ignition_thresholds}

Next, we discuss the impact of heating the as-milled powders to progressively higher temperatures to find the effect of heat removal on ignition thresholds. \cref{fig:wire_ignition_all_envs} confirms that ignition thresholds rise as heat is removed through annealing and we find that the degree of rise in ignition threshold is related to powder chemistry. \cref{fig:wire_ignition_air} shows that for the wire ignition tests performed in air, the relative rise in ignition threshold from as-milled to 100\% heat removed is inversely related to the Zr content of the powders. As available heat is removed, the powders' ability to generate heat via oxidation and/or nitridation increasingly controls the ignition threshold. 

The rise in ignition threshold is most drastic for 3Al:Zr, where, after 80\%  of the heat is removed, the powders are unable to ignite off the wire. This can be explained by the poor oxidation behavior of 3Al:Zr. Not only does 3Al:Zr demonstrate the lowest relative degree of mass gain from oxidation (compared to its theoretical maximum), it onset of oxidation is delayed the most. The poor oxidation behavior of 3Al:Zr is likely due to the more extensive formation of \ce{Al2O3} on the powders' exteriors. Despite \ce{Al2O3} being thermodynamically favored to form over \ce{ZrO2} at low temperatures \cite{balart2016grain}, the diffusivity of oxygen in \ce{Al2O3} is significantly lower than the diffusivity of oxygen into \ce{ZrO2} ($D_{\ce{Al2O3}}=4.7\times 10^{-28} \text{m}^2/\text{s}$ vs $D_{\ce{ZrO2}}=2.2\times 10^{-16} \text{m}^2/\text{s}$  respectively) \cite{ prot1996self, brossmann1999oxygen}. Thus, powders with greater Al content are expected to have slower oxidation, which is the dominant factor in controlling ignition after annealing.

Meanwhile, the rise in ignition threshold is the lowest for Al:3Zr. The DTA/TGA results in \ce{Ar} + \ce{O2} and \ce{Ar} + \ce{N2} show that Al:3Zr exhibits the most complete mass gain due to oxidation and nitridation after heating to 1000 °C. We attribute this to the greater volume fraction of Zr and the rapid diffusion of oxygen through \ce{ZrO2}, which facilitates fast and complete oxidation. The observed decrease in ignition threshold—from 55\% at 550 °C to 100\% at 1000 °C—for Al:3Zr is likely due to an increased presence of elemental Zr in the individual powders, rendering them richer in Zr than the nominal 1:3 atomic ratio, as supported by the Al-Zr phase diagram \cite{murray1992alzrphaseD}.

To investigate this hypothesis, we performed Rietveld refinement on Al:3Zr samples to estimate the volume fraction of elemental Zr after the powders were heated up to 1000 °C in an Ar atmosphere. A custom Python script was developed to quantify the Zr phase fraction by calculating its contribution to the total integrated area of the profile-fitted peaks using HighScore, following the methodology outlined in \cite{raguraman_machine_2024,raguraman2025simultaneous}. Additionally, XRD analysis of the Al:3Zr 750 °C sample was conducted to better understand the observed reduction in ignition threshold relative to the Al:3Zr 550 °C sample.

The estimated Zr volume fractions for Al:3Zr in its as-milled, 470 °C (36\%), 750 °C (100\%), and 1000 °C (100\%) states were 80.6\%, 71.9\%, 25.3\%, and 33.3\%, respectively. These results suggest that the drop in ignition threshold from Al:3Zr 550 °C (55\%) to Al:3Zr 1000 °C (100\%) is likely due to a relative increase in the amount of $\alpha$-Zr available for rapid oxidation.

Finally, Al:Zr exhibited similar behavior to Al:3Zr, and shows a plateau in ignition threshold once 75\% of the heat is removed. To understand why the ignition temperature plateaued, we likewise conducted XRD analysis on an additional sample of Al:Zr heated to 640 °C (76\%) in Ar to identify the starting phases present in the material. As shown in Supplementary Information Section 1.6, the XRD results indicate that Zr peaks are still present at 640 °C (76\%), but no free Zr remains after heating to 1000 °C. Although it might be expected that the absence of excess Zr in Al:Zr 1000 °C would lead to an increase in the ignition threshold, we observed a similar threshold to that of Al:Zr 640 °C (76\%). 

XRD analysis of the Al:Zr 1000 °C sample revealed the presence of the metastable \ce{Al2Zr} intermetallic, which may explain the stable ignition threshold. Oxidation studies by Paljevic \cite{paljevic1994high} demonstrated that \ce{Al2Zr} exhibits the earliest onset of oxidation among all Al-Zr intermetallics tested in dry oxygen at 0.3 K $\text{min}^{-1}$. Interestingly, \ce{Al2Zr} began oxidizing at even lower temperatures than $\alpha$-Zr with 1\% mass Al (357 °C versus 407 °C respectively) \cite{paljevic1994high}. As illustrated in \cref{thermal_analysis_mass_gain}(e), the Al:Zr 1000 °C sample tested in air showed an earlier onset of mass gain compared to as-milled Al:Zr, likely due to the rapid oxidation of the metastable \ce{Al2Zr} that is present. To demonstrate this, we cross-referenced the phases identified from the XRD results of the 3Al:Zr, Al:Zr, and Al:3Zr powders with the reported oxidation onset temperatures derived from Paljevic \cite{paljevic1994high}. The phase exhibiting the earliest oxidation onset in each case was highlighted on \cref{thermal_analysis_mass_gain}(d–f) using a dark red scatter point. 

For 3Al:Zr, \ce{Al3Zr} begins oxidizing around 527 °C, for Al:Zr \ce{Al2Zr} begins oxidizing at 357 °C while for Al:3Zr, $\alpha$-Zr oxidizes around 407 °C \cite{paljevic1994high}, aligning closely with the observed onset of mass gain for all three chemistries. It is important to note that the oxidation studies shown in \cite{paljevic1994high} were conducted using pure \ce{O2} rather than air and at slower heating rates than those in the present study, so the oxidation kinetics should not perfectly match the results reported here.

\cref{fig:wire_ignition_all_envs} presents the wire ignition thresholds for the three Al/Zr powders after annealing and testing in different gas environments. We observed that the presence of nitrogen gas generally lowers the ignition threshold in most cases. This trend was consistent across all annealed 3Al:Zr powders and most of the annealed Al:3Zr powders. When nitrogen reduced the ignition threshold, the difference between ignition in air and in \ce{Ar} + \ce{O2} was relatively small. This is likely because nitridation releases less heat than oxidation, as supported by the results in \cref{thermal_analysis_mass_gain}(a–c) and \cref{fig:thermal_analysis_summary_fig}.

Interestingly, none of the annealed 3Al:Zr or Al:3Zr powders ignited in \ce{Ar} + \ce{N2}, reinforcing that oxidation is critical for ignition in powders with reduced heat from intermixing or intermetallic reactions. However, nitrogen appeared to have a more pronounced effect on Al:Zr powders, particularly for Al:Zr 490 °C (50\%). Remarkably, Al:Zr 350 °C (21\%) was the only annealed powder that ignited in \ce{Ar} + \ce{N2}, likely due to its elevated heats of formation and nitridation—the highest of any tested chemistry at low temperatures (less than 450 °C). Notably, its ignition threshold (567.8 $\pm$ 28.35 °C) was the highest among all annealed powders that could still be measured with our setup, suggesting that oxidation is essential to sustain complete combustion.

It remains unclear why annealed Al:Zr powders exhibit lower ignition thresholds in \ce{Ar} + \ce{O2} than in air. One possible explanation is that nitridation competes with oxidation but contributes less heat overall. This effect may be more pronounced in Al:Zr due to its higher heats of intermixing, formation, and nitridation relative to 3Al:Zr and Al:3Zr.

\subsection{Combustion Properties}
\label{subsec:discussion_combustion}

The primary objective of this work was to evaluate whether annealing, which removes the intermetallic heat of formation, impacts the combustion properties of composite powders with identical composition and microstructure. To investigate this, we compared burn duration, combustion temperature, and microexplosion frequency between as-milled powders and their fully annealed counterparts. Interestingly, annealed powders that had formed intermetallic phases exhibited combustion properties comparable to, and in some cases exceeding those of the as-milled powders. This finding is significant because scalable manufacturing techniques, such as atomization, naturally produce powders with pre-formed intermetallic phases, rather than composite microstructures, as seen in ball milling. 

We also examined the influence of Zr content on combustion behavior across both as-milled and annealed powders. While as-milled powders generally exhibited consistent burn durations and temperature ranges regardless of Zr content, annealed powders showed greater variance with differences in Zr content. 
% Burn Duration

\subsubsection{Effect of Annealing}
\label{subsec:disc_annealing_combustion}

Annealed 3Al:Zr powders demonstrate longer burn durations, a substantial increase in average combustion temperatures, and a moderate decrease in microexplosion frequency compared to their milled counterparts. However, the annealed 3Al:Zr powders exhibited a much higher ignition threshold, and hence far fewer powders were analyzed in combustion, likely the small end of the size distribution. Thus, burn durations could be longer still. The longer duration is attributed in part to the recorded drop in microexplosion frequency, which is thought to limit the time of combustion. While statistical tests did not show a negative correlation between burn time and microexplosion for these powders, the differences in starting microstructures could change reaction kinetics to disfavor microexplosions. 

Previous studies have shown that microexplosions in Al/Zr composite powders occur when \ce{N2} bubbles that heterogeneously nucleate from high melting temperature impurities like \ce{HfN} or \ce{ZrN} rapidly expand at a critical growth rate \cite{wainwright2018gasenvs, wainwright2019bubbling}. The XRD results in \cref{XRD_panel} show that all powders initially have ZrC, which could function as heterogeneous nucleation sites for nitrogen bubbles since \ce{ZrC} melts around 3700 K \cite{manara2013zrc}. Annealing 3Al:Zr powders to form intermetallic \ce{Al3Zr} could impact the availability, size, and/or distribution of ZrC impurities, which may influence whether the particles microexplode. 

Heating Al:Zr powders to 1000 °C in Ar did not result in the same shifts in combustion behavior as seen for 3Al:Zr powders. Al:Zr demonstrated minimal changes between as-milled and annealed powders; there was a less than 100 K difference in average and average maximum temperature, and the average burn durations were nearly identical, both around 3 ms. There was also only a slight increase in microexplosion frequency after annealing. Meanwhile, annealed Al:3Zr powders exhibited slightly shorter average burn durations, significantly higher average temperatures (by approximately 500 K), and an increased probability of microexplosion compared to as-milled Al:3Zr. 

Since all three chemistries exhibited similar or higher combustion temperatures after annealing and the formation of Al-Zr intermetallic phases, the presence of distinct intermetallic phases appears to enhance, rather than degrade, combustion performance. Future studies will investigate whether atomized Al/Zr powders with equilibrium intermetallic phases exhibit combustion performance comparable to that of the annealed composite powders analyzed in this work and will examine whether specific intermetallic phases (e.g., \ce{Al2Zr}) perform better than others.

\subsubsection{Effect of Chemistry}
\label{subsec:disc_chemistry_combustion}

% Chemsitry -- burn duration
Next, we discuss the impact of increasing Zr content on combustion properties. Considering burn durations, we find that the majority of particles burned between 2 to 10 ms regardless of chemistry, which is comparable to values found in similar single particle experiments of various elemental and composite metal powders \cite{feng2020ignition, aly2015ignition}. The large standard deviations in burn durations can be attributed in part to the size and chemistry variations that are inherent to the ball-milling process. \cref{fig:combustion_results} shows that powder burn duration is dependent on whether durations are normalized by mean particle size. 

Since our setup does not allow direct measurement of individual particle sizes, we normalized burn durations using the mean particle diameter obtained from particle size analysis. In our initial comparison shown in \cref{fig:combustion_results}(b), we assumed a single proportionality constant across all chemistries and an exponent \(n\) = 2 based on theoretical models for pure Al powders. However, these assumptions are unlikely to hold for our composite systems. To still explore how increasing Zr content might influence combustion behavior burn duration, we instead assumed that the proportionality constant \(a\) remains the same across compositions, while systematically varying the exponent \(n\) between 1 and 2. A study on mechanically milled Al-Ti powders, with Ti content ranging from 10 to 25 at\%, found that using an exponent between 1 and 2 yielded good agreement by minimizing the normalized sum of squared errors of the radiation intensity profiles \cite{shoshin2006particle}.  It is likely that these ball-milled composite powders burn in a manner similar to Al/Ti powders, since Zr also burns in a condensed phase, like Ti. This approximation is imperfect, as both \(a\) and \(n\) are expected to vary with chemistry and we are still using the mean diameter found from particle size analysis, and not the mean particle size of the particles that combusted. 

Supplementary Information Figure SI 9 plots how the normalized as-milled burn durations vary as the exponent \(n\) in the burn duration power law is systematically adjusted between 0.5 and 2 with a constant value for \(a\). The as-milled 3Al:Zr powders burn for shorter durations compared to Al:Zr and Al:3Zr powders, which are close to one another. The same study reported that powders with higher Ti content exhibited longer burn durations \cite{shoshin2006particle}, which is consistent with the observation here that shows Al:Zr powders burn longer on average than 3Al:Zr powders once normalized. Al:Zr and Al:3Zr powders burned for similar average combustion times, and it is possible that adding a condensed-phase alloying element extends the burn duration up to a certain threshold, beyond which the composite's combustion behavior transitions to a predominantly condensed-phase regime. 

Contrary to expectations, we did not observe a negative correlation between microexplosion occurrence and burn duration for any of the as-milled powders. As a result, we cannot confidently conclude that variations in microexplosion frequency have a causal influence on burn duration. In future combustion studies, we plan to fabricate powders by Physical Vapor Deposition (PVD) over a broader range of Zr chemistries to isolate the effects of chemistry on burn duration.

% Chemistry -- temperature
Previous combustion studies of Al:Zr powders synthesized via PVD in air have measured combustion temperatures between 2300-3500 K \cite{wainwright2018gasenvs,alemohammad2020kilohertz}, and we found that all the Al/Zr composite powders burned across a similar range of temperatures. In general, the average combustion temperatures are similar to those found for micron-sized aluminum in air at ambient temperatures and pressures (2500-3100 K)\cite{badiola2011combustion}. The highest particle temperatures (3000-3500 K) are close to the maximum temperatures reported for single particle experiments of micron-sized Zr powders \cite{badiola2013combustion}. The average and average maximum combustion temperatures were highest for Al:Zr, then 3Al:Zr, and finally Al:3Zr which implies that there is no simple relationship between Zr content and average combustion temperatures. 

Still, we observed that the overall difference in average combustion temperatures for the as-milled powders is not vastly different; the averages vary by around 150-300 K. We suggest that the range of burn temperatures in these composite powders is tied to the Al boiling point at 2750 K and is bounded by the maximum temperatures observed in Zr combustion studies. However, an increase in Zr content did not result in a significant rise in the average burn temperature or burn duration, indicating that the potential benefits of incorporating substantial amounts of Zr into Al-based composite powders may be limited. 

% Chemistry -- microexplosion frequency
The as-milled powders frequently microexploded, ranging from 54.8\% for Al:3Zr to 66.3\% for Al:Zr. Therefore, we can conclude that for the as-milled particles that ignite, they have a high likelihood of microexploding. Once again, we observe that there is no clear trend in microexplosion frequency with respect to Zr content. From \cref{XRD_panel}, powders with higher Zr content show higher volume fractions of ZrC in as-milled powders, which potentially implies there are more available sites for bubble nucleation. X-ray phase contrast imaging of in-situ bubbling of Al-based composites has demonstrated that the rapid formation of many bubbles can prevent a single, large enough bubble from growing, thereby inhibiting microexplosions from occurring. Therefore, expecting Al:3Zr to show a higher microexplosion frequency than 3Al:Zr and Al:Zr may be too simplistic. 

The interplay between combustion temperature, the availability and distribution of bubble nucleation sites, and the chemistry affecting nitrogen dissolution into the molten particle likely governs the occurrence of microexplosions \cite{lange2022modelling}. Al-Zr composites consistently exhibit high microexplosion frequencies, suggesting that a range of factors determines these events. Future work will investigate the time-temperature traces for single particles to see if there are mechanistic differences between microexploding Al/Zr powders with varying Zr content. 

Looking ahead towards potential applications, our findings highlight a trade-off between ignition sensitivity and combustion performance in Al/Zr powders. Al-rich compositions may benefit from added Zr to reduce ignition thresholds, though excessive Zr yields diminishing returns in combustion output. An optimal Al–Zr ratio could be identified through broader combinatorial studies, including physical vapor deposition techniques that precisely control particle size and chemistry. 

%----------------------------------------------------------------------------------------------------------------------------------------
\section{Conclusions}

In this study, three ball-milled Al/Zr composite powders (3Al:Zr, Al:Zr, and Al:3Zr) were systematically annealed up to 1000 °C in argon to progressively reduce the heat available from intermetallic formation reactions. Hot-wire ignition experiments demonstrated that ignition thresholds are highly tunable, depending on the powder composition and degree of annealing. Considering the as-milled powders, Al:Zr had a significantly lowered ignition threshold (354 °C) compared to 3Al:Zr and Al:3Zr powders (434 °C and 444 °C, respectively), which is due to Al:Zr demonstrating faster initial heat release from atomic intermixing. It was shown that for all powders, ignition was driven by the exothermic heat release from intermixing and intermetallic formation, matching the results of prior studies.

As the powders were annealed, ignition thresholds rose sharply for 3Al:Zr which failed to ignite off the hot wire after 80\% of its available intermetallic heat was removed. In contrast, Al:Zr's ignition threshold plateaued at 520–540 °C after losing ~50\% of its available heat, while Al:3Zr showed only a slight increase after annealing to 1000 °C. Thus, we find that as the heat from intermixing and intermetallic reactions is removed, oxidation becomes the dominant ignition driver; this is particularly true for Zr-rich powders which oxidize more readily. The wire ignition results for annealed powders were corroborated by TGA data performed at slow heating rates, which showed that powders with more Zr had better oxidation behavior. 

The combustion experiments showed that as-milled powders burned over similar average temperatures (2450 K to 2750 K), which leads us to suggest that the combustion temperature for these powders is tied to the Al boiling point. Although there was an increase in the normalized burn duration from as-milled 3Al:Zr to Al:Zr, the as-milled Al:3Zr powder showed a normalized burn duration similar to that of Al:Zr, but also burned at a lower average combustion temperature. Therefore, beyond a certain Zr content, further increases do not appear to significantly improve combustion performance. Given the practical constraints associated with Zr—such as cost and limited availability—identifying an optimal Al/Zr ratio is essential for balancing performance and feasibility.

Annealed powders burned at significantly higher average temperatures (500–600 K) for 3Al:Zr and Al:3Zr compared to their as-milled counterparts, while annealed Al:Zr powders burned at nearly the same average temperature as their as-milled counterparts. Meanwhile, the average burn durations for as-milled versus annealed powders remained relatively constant. The cause of the large changes in average combustion temperature for 3Al:Zr and Al:3Zr powders remains unclear, but the observed increase in burn temperature with annealing is encouraging. These results suggest that annealed powders, despite containing pre-formed intermetallics, can match or even surpass the combustion performance of their as-milled counterparts. This finding supports the potential of using atomized Al/Zr powders with pre-formed intermetallics as a more cost-effective and scalable alternative to ball-milled composites for fabricating reactive powders.

Finally, microexplosion frequency was successfully assessed using a novel CNN-based detection pipeline, which confirmed frequent microexplosions across all compositions and conditions (>46\%). The fact that no simple correlations were found between microexplosion frequency and Zr content or annealing suggests that the occurrence of microexplosions depends on a complex interplay of combustion temperature, bubble nucleation sites, powder chemistry, and the presence of pre-formed intermetallics. Future work is underway using X-ray Phase Contrast Imaging to investigate the nucleation and growth of bubbles during combustion and elucidate the mechanisms responsible for microexplosions in ball-milled composites with different metal additives.

%----------------------------------------------------------------------------------------------------------------------------------------

\section{Declaration of competing interest}
The authors declare that they have no known competing financial interests or personal affiliations that might have influenced the findings presented in this paper.
%----------------------------------------------------------------------------------------------------------------------------------------

\section{Acknowledgements}
This research was supported by the Department of Defense, Defense Threat Reduction Agency (DTRA) under the MSEE URA (Grant HDTRA1-20-2-0001). Michael R. Flickinger, Colin Goodman, Megan Bokhoor, Rami Knio, Michael Kruppa, Mark A. Foster, and Timothy P. Weihs received support through the MSEE URA. Sreenivas Raguraman was partially supported by the National Science Foundation under Grant DMR \#2320355. Amee L. Polk was supported by the Department of Defense Science, Mathematics, and Research for Transformation (SMART) Scholarship-for-Service, and the U.S. Army's Chemical Biological Advanced Materials and Manufacturing Science Program at the Combat Capabilities Development Command Chemical Biological Center.

The authors gratefully acknowledge the Materials Characterization and Processing Facility and the Hopkins Extreme Materials Institute for providing access to characterization and sample preparation facilities. The authors would like to thank Dr. John Fite, Dr. Jesse Grant, and Preetom Borah for their helpful insights and discussions during the course of this work.  
%----------------------------------------------------------------------------------------------------------------------------------------
%\clearpage
\bibliography{references}
\bibliographystyle{elsarticle-num} 

%----------------------------------------------------------------------------------------------------------------------------------------

\end{document}